\documentclass[acmtog]{acmart}
\AtBeginDocument{%
  \providecommand\BibTeX{{%
    \normalfont B\kern-0.5em{\scshape i\kern-0.25em b}\kern-0.8em\TeX}}}

\usepackage{wrapfig}
\usepackage{caption}
\usepackage{subcaption}
\usepackage{placeins}
\usepackage{float}
\usepackage{stfloats}

\copyrightyear{2023} 
\acmYear{2023} 
\setcopyright{acmlicensed}\acmConference[FAccT '23]{2023 ACM Conference on Fairness, Accountability, and Transparency}{June 12--15, 2023}{Chicago, IL, USA}
\acmBooktitle{2023 ACM Conference on Fairness, Accountability, and Transparency (FAccT '23), June 12--15, 2023, Chicago, IL, USA}
\acmPrice{15.00}
\acmDOI{10.1145/3593013.3594020}
\acmISBN{979-8-4007-0192-4/23/06}

\newcommand\numTotalImages{{24,803,854}~} %

\newcommand\multipleHighesttoLowestExposure{{almost 20}~}
\newcommand\numPhaseOneImages{{3,987,835}~}
\newcommand\numPhaseTwoImages{{20,816,019}~}
\newcommand\numImagesAnnotated{{15,250}~}

\newcommand\percentHoursCoveredPhaseTwo{{100}\%~}

\newcommand\meanImagesPerCBG{{168.2}~}
\newcommand\percentCBGWithAtLeastOneImage{{99.6}\%~}

\newcommand\scaleAIPoliceImages{{1,088}~}
\newcommand\SSLImageCount{{450,000}~}
\newcommand\trainSetSize{{9,449}~}
\newcommand\finalTrainSetPositives{{4,748}~}
\newcommand\finalTrainSetNegatives{{4,701}~}

\newcommand\numImagesAboveThreshold{{233,596}~}

\begin{document}

\title{Detecting disparities in police deployments using dashcam data}

\author{Matt Franchi}
\email{mwf62@cornell.edu}

\affiliation{%
  \institution{Cornell Tech}
  \streetaddress{2 West Loop Rd}
  \city{New York}
  \state{New York}
  \country{USA}
  \postcode{10044}
}

\author{J.D. Zamfirescu-Pereira}
\affiliation{%
  \institution{University of California - Berkeley}
  \streetaddress{}
  \city{Berkeley}
  \country{USA}}
\email{zamfi@berkeley.edu}

\author{Wendy Ju}
\affiliation{%
 \institution{Jacobs Technion-Cornell Institute, Cornell Tech}
 \city{New York}
 \country{USA}
 }

\author{Emma Pierson}
\affiliation{%
  \institution{Jacobs Technion-Cornell Institute, Cornell Tech}
  \city{New York}
  \country{USA}
  \email{emma.pierson@cornell.edu}
}

\renewcommand{\shortauthors}{Franchi et al.}

\begin{abstract}
Large-scale policing data is vital for detecting inequity in police behavior and policing algorithms. However, one important type of policing data remains largely unavailable within the United States: aggregated \emph{police deployment data} capturing which neighborhoods have the heaviest police presences. Here we show that disparities in police deployment levels can be quantified by detecting police vehicles in dashcam images of public street scenes. Using a dataset of \numTotalImages dashcam images from rideshare drivers in New York City, we find that police vehicles can be detected with high accuracy (average precision 0.82, AUC 0.99) and identify \numImagesAboveThreshold images which contain police vehicles. There is substantial inequality across neighborhoods in police vehicle deployment levels. The neighborhood with the highest deployment levels has \multipleHighesttoLowestExposure times higher levels than the neighborhood with the lowest. Two strikingly different types of areas experience high police vehicle deployments --- 1) dense, higher-income, commercial areas and 2) lower-income neighborhoods with higher proportions of Black and Hispanic residents. We discuss the implications of these disparities for policing equity and for algorithms trained on policing data.
\end{abstract}

\keywords{dashcam, object detection, policing}

\maketitle

\section{Introduction}

How do citizens hold the police to account, ensuring that they are equitably protecting those they serve? A vital tool in the struggle for police accountability has been \emph{data} shedding light on police activity. Statistical analysis of police data has been used to document disparities in police stops, searches, use of force, and respect~\cite{gelman2007analysis,goel2016precinct,goncalves2021few,pierson2020large,pierson2018fast,anwar2006alternative,voigt2017language,fryer2019empirical,simoiu2017problem}. These statistical analyses have been instrumental to driving policy change and police reform: examples include a reduction in random police searches in Los Angeles following a statistical investigation documenting racial bias~\cite{la_times_policing_1,la_times_policing_2} and a consent decree in Chicago following a Department of Justice investigation documenting numerous rights violations~\cite{doj_investigation_chicago_police_dept}. %
In recent years, policing datasets have become more widely available to the public~\cite{pierson2020large, washington_post_police_shootings,police_scorecard}, allowing activists, policymakers, and researchers to examine a range of policing behavior, including stops, searches, and arrests.

However, there is still one type of data that remains largely unavailable to the public within the United States: information on where the police are deployed in the first place. This data is crucial for several reasons. First, statistical analyses of police deployments have revealed inefficiencies or inequities~\cite{nytimes_bronx_article,nytimes_bronx_article_2} in which some areas have disproportionate police deployments. Second, disparities in police deployments create biases in downstream outcomes like arrests. If an area has a heavier police presence, crimes are more likely to result in arrests. Thus, more heavily policed neighborhoods may appear to be more prone to crime, when they are simply more prone to \emph{observed crime}; similarly, residents of these neighborhoods may appear to be more likely to commit a crime, when in fact they are simply more likely to be arrested for it. This bias has been observed in practice: for example, African Americans are far more likely to be \emph{arrested} for marijuana use than are white Americans, even though surveys of drug \emph{use} do not show the same racial disparities~\cite{ramchand2006racial}. This bias could also propagate into \emph{algorithms} used in policing and criminal justice --- including pretrial risk assessments~\cite{laura2016public,corbett2017algorithmic,corbett2016computer} and predictive policing algorithms~\cite{brookingspredictivepolicing,lum2016predict,baltimoresunpredictivepolicing,richardson2019dirty,selbst2017disparate} --- since these algorithms can use arrests, or outcomes downstream from arrests like convictions, as input. %

Deployment data are thus essential for monitoring disparities or inefficiencies in policing and detecting potential biases in algorithms. Even \emph{aggregated} deployment data --- grouped by neighborhood or demographic --- would be useful towards this end. In spite of this, police deployment data remains largely unavailable to the public within the United States. Addressing this need, we introduce a novel methodology for monitoring police deployments: detecting police vehicles in dashcam images of public street scenes. Using a dataset of more than 24 million dashcam images collected throughout New York City in 2020, we annotate a training dataset of \trainSetSize images for presence of police vehicles, and train a deep learning model to identify police vehicles with high accuracy (average precision 0.82; AUC 0.99). We use this model to identify \numImagesAboveThreshold dashcam images which contain police vehicles. We develop a framework for analyzing inequality in police deployment levels across neighborhoods  --- that is, the probability an image within a given neighborhood contains a police vehicle --- which compensates for several data and model biases. Our analysis reveals substantial disparities in police deployments: the neighborhood with the highest police deployments has \multipleHighesttoLowestExposure times higher deployments than the neighborhood with lowest deployments. Two very different types of areas experience high police deployments: dense, high-income, commercial areas; and areas with higher proportions of Black and Hispanic residents and lower incomes. We discuss the implications of these disparities for policing and algorithmic equity. We also discuss the residual potential biases in the data which cannot be removed by our extensive debiasing pipeline, and argue that this fact suggests that the police should simply release more reliable and straightforward aggregated deployment data.  %

\section{Related Work}

\subsection{Auditing policing practices}

An extensive academic literature uses large-scale policing data to statistically audit policing for efficiency and equity~\cite{gelman2007analysis,goel2016precinct,goncalves2021few,pierson2020large,pierson2018fast,anwar2006alternative,voigt2017language,fryer2019empirical,lum2016predict,simoiu2017problem,ridgeway2009doubly,grogger2006testing,warren2006driving,ang2021effects}. (Here and throughout the paper, we focus our discussion on policing within the United States because policing practices are highly heterogeneous across countries and our data is from New York City.) The existing academic literature largely analyzes outcomes \emph{downstream} of police deployment patterns: that is, it studies not where police \emph{go}, but what they \emph{do once they get there}. %
For example, many papers~\cite{gelman2007analysis,goel2016precinct,pierson2020large,pierson2018fast,ridgeway2009doubly,grogger2006testing,warren2006driving} quantify disparities in how likely the police are to stop drivers of different races. Work has also studied post-stop outcomes: for example, ~\cite{goncalves2021few} studies racial disparities in whether police grant leniency to drivers pulled over for speeding; ~\cite{pierson2020large,pierson2018fast,anwar2006alternative,simoiu2017problem,jung2018omitted} study racial disparities in police searches; and ~\cite{voigt2017language} studies racial disparities in how respectful officers are to drivers of different races. Work also quantifies police misconduct~\cite{headley2020effect}; use of force~\cite{fryer2019empirical}; police killings~\cite{edwards2019risk}; and downstream effects of police violence~\cite{ang2021effects}.

Outside of academia, there have also been extensive journalistic, activist, and legal efforts to quantify disparities in policing. Journalistic efforts include an investigation by the Los Angeles Times~\cite{la_times_policing_1,chang2019stop} which documented racial bias in policing practices, ultimately resulting in a curtailment of random vehicle searches in an effort to reduce racial bias~\cite{la_times_policing_2}. Closer to our own work, ~\citet{nytimes_bronx_article} analyzed confidential data on detective deployments within New York City and found racial and socioeconomic inequality in whether neighborhoods had enough homicide detectives to meet their needs. This report was followed by calls for increased transparency and more equitable deployments~\cite{nytimes_bronx_article_2}, testifying to the importance of police deployment data. Activists have also compiled large datasets on police activity~\cite{police_scorecard}. Finally, many legal efforts against the police have leveraged large-scale policing data~\cite{Floyd,doj_investigation_chicago_police_dept}. 

Other efforts have quantified disparities in policing using datasets other than large-scale data from the police themselves. For example, journalists have analyzed footage of individual police encounters~\cite{new_york_times_visual_investigations}; researchers have also interviewed individuals about their experiences with police~\cite{epp2014pulled} and conducted analyses of survey data on experiences with police~\cite{davis2018contacts}.

\subsection{Effect of deployment disparities on algorithmic bias}%

Many algorithms used in policing and criminal justice use data from outcomes downstream from police deployments, including arrests and convictions. Examples include pretrial risk assessments~\cite{laura2016public,corbett2017algorithmic,corbett2016computer} and predictive policing algorithms~\cite{brookingspredictivepolicing,lum2016predict,baltimoresunpredictivepolicing,richardson2019dirty,selbst2017disparate}.\footnote{We note that the implementation details of many of these algorithms remain murky, and not all of them rely on arrest or conviction data.} Thus, disparities in police deployments can bias the data these algorithms use~\cite{corbett2017algorithmic,brookingspredictivepolicing,lum2016predict,baltimoresunpredictivepolicing,richardson2019dirty,selbst2017disparate}. For example, ~\citet{lum2016predict} argue that disparities in police deployments could result in increasingly severe biases over time in predictive policing algorithms trained on arrest data. In their model, heavy police deployments in areas result in higher arrest levels at the same crime levels; consequently, the algorithm learns that areas with heavy police deployments are riskier than they really are, and recommends that even more police be deployed there; this further biases the arrest rates which are fed back into the algorithm, exacerbating inequality over time. 

\subsection{Analysis of public street scene datasets}

Streetview imagery datasets -- for example, Google Street View, Microsoft Bing Maps, Baidu Total View, and Tencent Street View -- are increasingly used for urban analytics. Research has used such datasets to count trees \cite{thirlwell2020big, 8930566}, utility poles \cite{krylov2018automatic}, traffic signs \cite{campbell2019detecting, 10.1145/3180496.3180638}, bike racks \cite{maddalena2020mapping} and manholes \cite{vishnani2020manhole}. Streetview data has also been used for computational social science purposes: for example, estimating the demographic makeup of neighborhoods~\cite{gebru2017using} or quantifying human political stereotyping~\cite{zamfirescu2022trucks}.

More recently, \emph{dashcam datasets} collected by Automotive Original Equipment Manufacturers (OEMs) and add-on manufacturers such as Nexar or Mobileye that provide cloud-connected dashcam products have enabled street view image datasets to be collected at a greater spatial scale and temporal density \cite{yu2020bdd100k} (in contrast to Streetview data, which may be collected at much lower temporal density). This type of data has been used to assess pavement quality and classify road surface types \cite{dadashova2021detecting}, detect lanes or curbs \cite{zhou2020lacnet}, or detect potholes and other defects \cite{riedl2020importance}. 

Dashcam data has a unique combination of properties that make it apt for the automated detection of objects in dense, feature-rich environments. Dashcam data is frame-by-frame, inexpensive to gather, not fixed in terms of camera perspective, and capable of depicting pedestrians, other vehicles, and small objects in detail. %
Dashcam datasets have been used to characterize when and where people were out in New York City during different phases of curfew and social distancing policy~\cite{ju2020tracking, chowdhury2021towards, chowdhury2021tracking}. This work points to the possibility of using the density of the data from dashcam datasets to look at more dynamic aspects of urban environments--the number of people and cars, the density of traffic, and the prevalent activities~\cite{chowdhury2021tracking}.

Closest to our own work in this genre is ~\citet{sheng_surveilling_2021}, which uses Streetview data to quantify how the density of surveillance cameras varies across and within cities. However, our work differs in important ways: most notably, we study police deployment, not surveillance cameras. Overall, the works are complementary: ~\citet{sheng_surveilling_2021} analyzes surveillance cameras across 16 cities at lower spatial and temporal resolution, and we analyze police deployments across a single city at high resolution. %

\section{Dataset}

We estimate the total number and spatial distribution of police vehicles visible in public street scenes in New York City using a large dataset of dashcam images collected throughout 2020. In this section, we describe the dashcam dataset. In the next section, we describe our pipeline for analyzing it. 

\subsection{Dataset details}
Our dataset is provided by Nexar, a company which provides ride-sharing (Uber, Lyft, NYC Taxi, etc) drivers with dashboard cameras (dashcams) to record their drives (which can be useful if, for example, an accident occurs). The dataset consists of 24,803,854 images taken throughout the five boroughs\footnote{New York City is divided into five areas, referred to as boroughs: Queens, Brooklyn, Manhattan, the Bronx, and Staten Island.} of New York City between March 4 2020 and November 15 2020. Each image is 1280 x 720 pixels. We remove a small number of duplicate images (less than 0.01\% of the overall dataset) with identical latitudes, longitudes, and timestamps.

\subsection{Geographic and temporal coverage}
Data was provided to us by Nexar in two phases. Phase 1 consists of \numPhaseOneImages images sampled prior to September 1 2020, and is extremely geographically and temporally skewed. Geographically, it is concentrated within the boroughs of Manhattan and Brooklyn, and does not contain data from the boroughs of Staten Island, Queens, and the Bronx at all; temporally, it overrepresents data from Thursday nights. Phase 2, which constitutes the majority of the dataset, consists of \numPhaseTwoImages images sampled after October 4 2020, and is much more geographically and temporally representative: it is sampled at all times of the day, on all days of the week, and also covers the entire geographic area of New York City. 

\begin{figure}
  \includegraphics[width=0.5\textwidth]{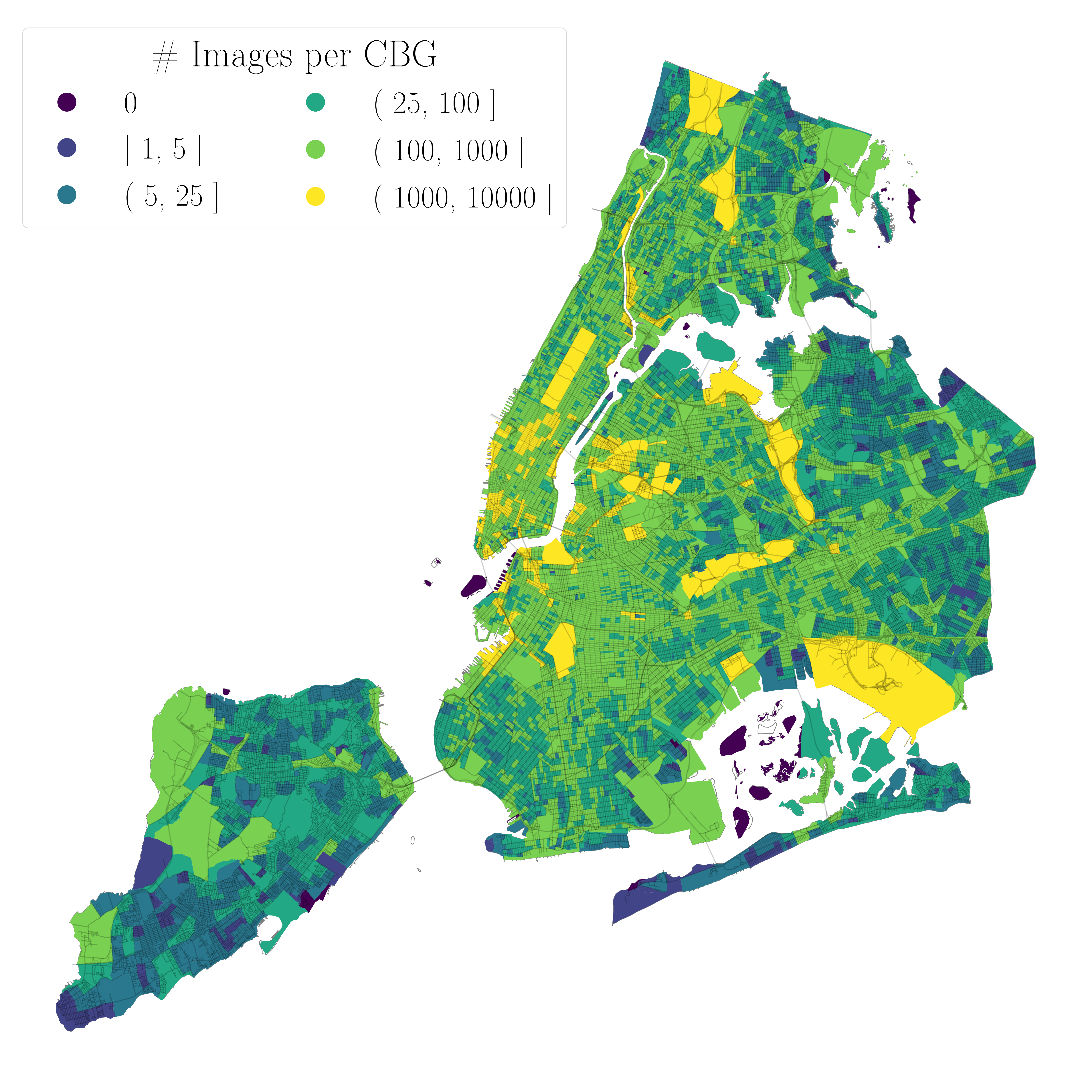}
  \caption{Number of images per Census Block Group in the Phase 2 dataset we use in our primary analysis of disparities.} %
  \label{fig:coverage}
\end{figure}

Because Phase 2 is much more representative than Phase 1, we conduct our primary analysis of disparities using only data from Phase 2. We additionally conduct numerous validations and bias corrections, described in \S \ref{sec:conceptual_framework}, to compensate for non-representative sampling in the dataset. Geographic and temporal coverage during the Phase 2 period is very good. Specifically, \percentHoursCoveredPhaseTwo of hours during the Phase 2 period are covered; \percentCBGWithAtLeastOneImage of Census Block Groups (CBGs)\footnote{Census Block Groups are Census areas consisting of a couple hundred to a couple thousand people, and are the finest geographic resolution at which we conduct our analysis. New York City has more than 6,000 Census Block Groups.} have at least one image, with a mean of \meanImagesPerCBG images per CBG; 88\% of roads contained within the borders of New York City are covered by at least one image, using data from OSMNX \cite{boeing_osmnx_2017}.  Figure \ref{fig:coverage} summarizes geographic data availability; Figure S1
summarizes temporal data availability.

\subsection{Ethical considerations} 

Our use of this dataset was deemed not human subjects research by our university's Institutional Review Board. Nonetheless, we carefully considered the costs and benefits of this research prior to embarking on it. A key concern for this type of research is that the "subjects"--police vehicles, other vehicles, and other bystanders--do not consent to the initial data collection nor its subsequent use in this research. However, under American Psychological Association Ethics guidelines \cite{american2002ethical}, informed consent procedures can be waived when permitted by law, federal and institutional regulations, or when the research would not reasonably be expected to distress or harm participants and involves naturalistic observations or archival research for which disclosure of responses would not place participants at risk of criminal or civil liability or damage their financial standing, employability or reputation, and for which confidentiality is protected. These provisions are the primary reason why naturalistic observation in public settings does not normally require informed consent~\cite{hill2011human}. Ultimately, we felt that the potential benefits of this work --- namely, quantifying disparities in police deployments to enable more equitable policing --- more than offset the costs. This conclusion is consistent with the numerous research works~\cite{sheng_surveilling_2021,gebru2017using,zamfirescu2022trucks,ju2020tracking, chowdhury2021towards, chowdhury2021tracking} which have used similar datasets capturing public street scenes for research. However, we do believe that it would benefit the public to have broader awareness that data from instruments such as dashcams are now being aggregated on a large-scale basis.

\section{Statistical analysis} 

In this section, we describe our procedure for assessing disparities in police deployments. First, we describe the mathematical framework to estimate disparities in police deployments while compensating for potential data and model biases (\S \ref{sec:conceptual_framework}). We then provide detail on the deep learning model, which is a key ingredient to this framework, as it detects police vehicles from dashcam images (\S \ref{sec:deep_learning_model}).

\subsection{Mathematical framework}
\label{sec:conceptual_framework}

Our goal is to assess disparities in police vehicle deployments while mitigating the effects of potential model or dataset biases. Here we introduce our mathematical framework for doing this. Before describing the details of the framework, the high-level intuition is that we are reweighting the Nexar data sample, which is sampled from a non-representative set of locations, so that it matches the locations where different demographic groups actually live. For example, to calculate the police deployment levels that Asian residents of New York City experience, we reweight the original data sample to upsample neighborhoods with larger Asian populations. 

More formally, let $A$ denote the demographic variable --- for example, racial group --- over which we wish to assess disparities in police deployment, and let $a$ denote specific values of this variable. Our goal is to estimate the probability a person of group $a$ has a police vehicle visible when they are on a street in their home Census area: in other words, to estimate $\text{Pr}(y=1|A=a)$, where $y$ indicates whether there is a police vehicle visible. Letting $C$ denote Census area, and summing over Census areas, we can write: 
\begin{equation}\label{eq:sum_over_census}
\small
\begin{split}
 \text{Pr}(y=1|A=a) &= \sum_{c} \text{Pr}(C=c|A=a) \cdot \text{Pr}(y=1|C=c, A = a) \\ %
 &= \sum_{c} \text{Pr}(C=c|A=a) \cdot \text{Pr}(y=1|C=c)
\end{split}
\end{equation}

where the second line reflects our assumption that conditional on Census area, the probability a police vehicle is visible is constant across groups: that is, one group within a Census area is no more likely to observe a police vehicle than another. We use fine-grained Census areas throughout our analysis --- Census Block Groups --- to render this assumption reasonable.\footnote{We use data from the 2020 American Community Survey.} 

We estimate the first term in the product, $\text{Pr}(C=c|A=a)$, as the fraction of people of group $a$ who live in Census area $c$. Let $N_{ca}$ denote the number of people of group $a$ living in Census area $c$. Then, letting $\widehat{\text{Pr}}(C=c|A=a)$ denote our estimate of the true probability $\text{Pr}(C=c|A=a)$, our estimator is: 
\begin{equation}\label{eq:reweighting_data}
\begin{split}
\widehat{\text{Pr}}(C=c|A=a) = \frac{N_{ca}}{\sum_{c} N_{ca}}
\end{split}
\end{equation}
The other term we must estimate to compute Equation \ref{eq:sum_over_census} is $\text{Pr}(y=1|C=c)$. Let $\hat y = 1$ if our police vehicle detection model detects a police vehicle in an image and $\hat y = 0$ otherwise. Then our estimator is:
\begin{equation}\label{eq:checking_model}
\begin{split}
\widehat{\text{Pr}}(y=1|C=c) =\hspace{1mm}  &\widehat{\text{Pr}}(\hat y=1|C=c) \cdot \widehat{\text{Pr}}(y=1|\hat y=1)\text{ } + \\ 
                      &\widehat{\text{Pr}}(\hat y=0|C=c) \cdot  \widehat{\text{Pr}}(y=1|\hat y=0)
\end{split}
\end{equation}
In other words, we estimate the fraction of images which truly have police vehicles in a Census area by taking the fraction classified positive multiplied by the estimated classifier precision $\widehat{\text{Pr}}(y=1|\hat y=1)$, plus the fraction classified negative multiplied by the estimated classifier false omission rate $\widehat{\text{Pr}}(y=1|\hat y=0)$. This procedure compensates for imperfect classifier performance. This procedure will be invalid if classifier performance --- i.e., $\text{Pr}(y=1|\hat y=1)$ and $\text{Pr}(y=1|\hat y=0)$ --- does not remain constant across Census areas, and in particular if it varies by demographic group $a$. Given the extensive literature illustrating that classifier performance can vary by by demographic group~\cite{buolamwini2018gender,koenecke2020racial,chen2021ethical,corbett2018measure,wilson2019predictive}, this is important to verify, and in \S \ref{sec:model_validation} we do so. %

Overall, our estimation procedure compensates for two types of potential bias. Equation \ref{eq:reweighting_data} compensates for a \emph{data bias}, reweighting the Nexar dataset (which is sampled from a set of locations which does not necessarily match the population distribution; Figure ~\ref{fig:coverage}) to match the population distribution of demographic subgroups. This is conceptually similar to inverse propensity weighting procedures~\cite{austin2015moving} which are used to compensate for non-representative data in other settings. Equation \ref{eq:checking_model} compensates for imperfect model performance, and allows us to check that model performance is unbiased (i.e., calibrated) across demographic subgroups. 

We note that there are other biases which could prevent us from perfectly estimating disparities in police deployments. In the Discussion section, we discuss two potential biases in more detail. First, the sample of Nexar images we have \emph{within} each Census area $c$ may be biased, in the sense that it may not capture the true probability a resident of that Census area will see a police vehicle at any given time when they are out on the road. For example, if vehicles are prohibited from driving near protest areas, which also have larger police presences, we will not have images of large police presences near protests. It is not possible to correct for this bias with the data we have because 1) the true distribution may differ from the Nexar sampling distribution along unobservable dimensions which we cannot reweight along and 2) we may simply have no Nexar images in some regions of the true distribution (e.g. if all vehicles are banned near protests). A second potential bias is that police vehicles represent only a subset of overall police activity: for example, they do not capture officers on foot. We return to both these points below.

\subsection{Deep learning model} 
\label{sec:deep_learning_model}

We train a deep learning model to identify police vehicles with high accuracy. Our model training pipeline consists of three steps. First, we annotate a large dataset of images  for presence of police vehicles (\S \ref{sec:police_vehicles_dataset}); second, we train a police vehicle detection model on this dataset (\S \ref{sec:model_training}); finally, we use a held-out dataset to verify that the model achieves high accuracy on the dataset overall and is not biased with respect to demographic groups  (\S \ref{sec:model_validation}). %

\subsubsection{Data annotation} 
\label{sec:police_vehicles_dataset}

We begin by creating training, validation, and test datasets for our police vehicle detection model. Following standard machine learning practice, we use the training set to optimize the parameters of each machine learning model; the validation set to select the model which yields the best performance; and the test dataset to measure the performance of the chosen model. We draw all training images from a random 10\% of the dataset, and reserve the remaining 90\% for validating the model and analyzing disparities. 

A challenge in creating the training set is that the dataset is imbalanced --- the vast majority of images do not contain police vehicles --- and highly imbalanced datasets can produce inferior model performance~\cite{japkowicz2002class}. To mitigate this issue, we construct a more balanced training set by sampling images near police stations, which are more likely to contain police vehicles: specifically, we filter for images within 0.25 miles of one of NYC's 77 police precinct stations, which raises the proportion of images with police vehicles from about 1\% to roughly 7\%. The risk of this training strategy is that it potentially introduces distribution shift~\cite{koh2021wilds}, which can also harm model performance, because images near police stations may differ from the overall image distribution; below, we describe how we check for this concern. 

We collect annotations for \numImagesAnnotated images using Scale.AI \cite{scaleai}, an outsourcing platform for annotations similar to Amazon Mechanical Turk. Each image is annotated with rectangular bounding boxes to identify police vehicles. We utilize evaluation tasks (images pre-verified with ground truth annotations by us) on the platform to disqualify inaccurate labelers. Every image labeled by a Scale.AI annotator is then submitted to a review phase, where a separate, high-scoring labeler checks each image for accuracy. In choosing which types of NYPD vehicles to have annotators label, we account for the fact that the NYPD has several divisions, including the prisoner transport, building maintenance, and museum units \cite{noauthor_nypd_nodate}, 
that are not involved in active policing. We instruct annotators to only label NYPD sedans, SUVs, compacts, and trucks (but not buses or 4x4 trucks) to mitigate this. In total, \scaleAIPoliceImages images from Scale.AI have police vehicles; we add these to the training set.

To further increase the size of the training dataset, we use the model trained on the Scale.AI annotations to select images which it predicts have police vehicles (at a confidence threshold of 0.9). Specifically, we apply the model to \SSLImageCount images selected at random from the 10\% training sample (not just images near police stations); examine all images passing the 0.9 confidence threshold and manually verify whether these images do, in fact, contain police vehicles, and add the images with verified labels to the train set. The final training set consists of \finalTrainSetPositives images with police vehicles and \finalTrainSetNegatives images without police vehicles. %

In contrast to the training set, which oversamples images near police stations, the validation and test splits are both random samples of 20,000 images from the overall dataset. A random sample is essential because it provides us with an unbiased measure of performance, which our estimation framework requires (\S \ref{sec:conceptual_framework}). To annotate the validation and test sets, we use Scale.AI to collect annotations and manually audit all images labeled as positive for correctness. The validation set includes 247 positive images (i.e., which contain police vehicles) out of 20,000, and the test set includes 239 positive images out of 20,000. 

\subsubsection{Model training} 
\label{sec:model_training} 

We frame our police vehicle detection problem as a rectangular object detection problem. We fine-tune models from the widely-used You Only Look Once (YOLO) suite of object detection models~\cite{wang2022yolov7} which have been pretrained on the MS-COCO dataset to detect object classes including vehicles~\cite{fleet_computer_2014}. We compare performance from a range of model configurations (for example, we compare performance from model variants YOLO5X~\cite{yolov5} and YOLO7-E6E~\cite{wang2022yolov7}, and experiment with hyperparameter evolution). We use average precision on the validation set to select the model which yields optimal performance.\footnote{Because each image can potentially contain multiple police vehicles and multiple model predictions, to compute average precision, we define an overall model prediction for each image as the maximum model prediction for any object detection (or as 0 if no objects are detected). We define the true label for each image as 1 if it contains any police vehicles, and 0 otherwise. We do this because it allows us to compute precision and recall statistics at the per-image level, not the per-object level, and our primary analysis is at the per-image level.} Our final model is trained using the default hyperparameter configuration for YOLOv7. To choose a threshold for positive classification, we use the threshold (0.77) which maximizes the F-score on the validation set.

\subsubsection{Model validation} 
\label{sec:model_validation} 

\begin{figure}
    \begin{center} 
        \includegraphics[trim=82px 0 0 0, clip,width=0.53\textwidth]{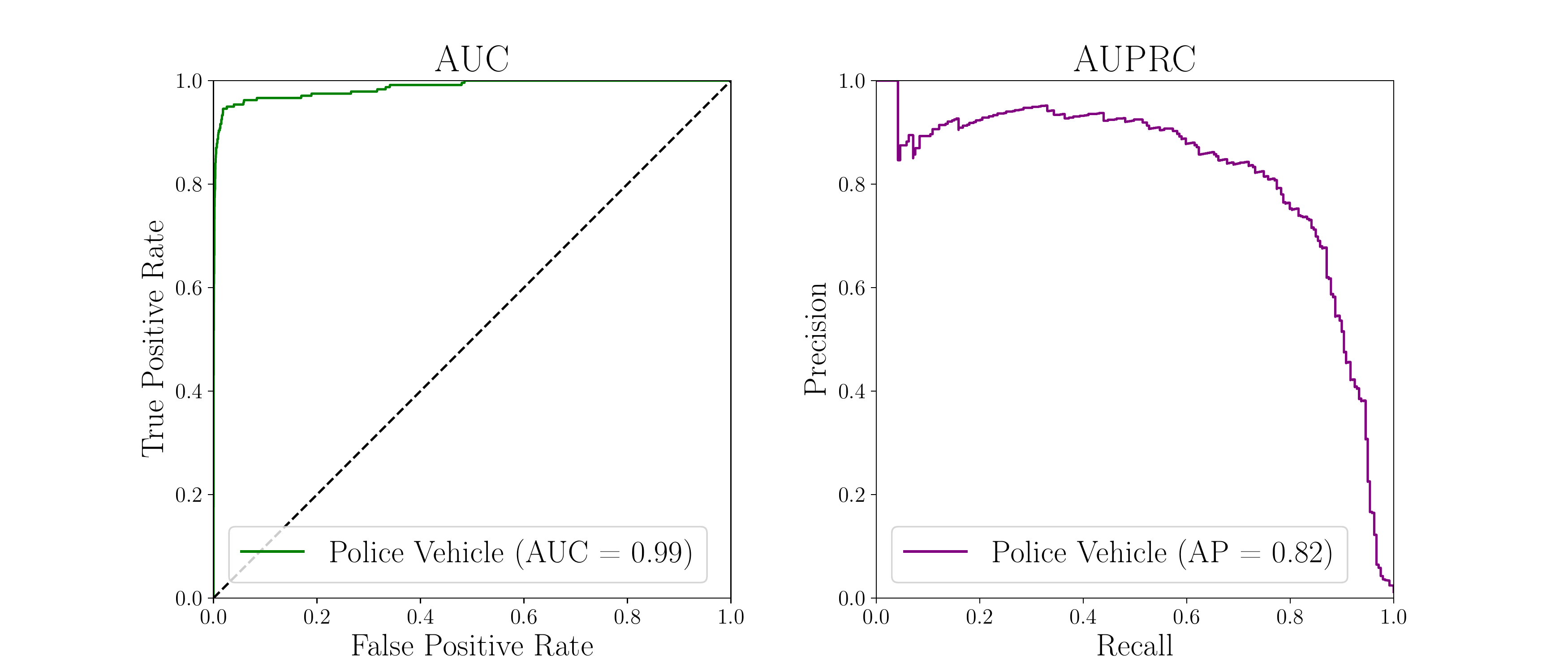}
        \caption{Receiver Operating Characteristic (ROC) and Precision-Recall curves for our final model, evaluated on the test set.}
        \label{tab:test_set_performance_figure}
    \end{center}
\end{figure}

\begin{figure*}
    \centering
    \begin{subfigure}[b]{0.35\textwidth}
         \centering
         \begin{tabular}{ll}
            \includegraphics[width=.5\textwidth]{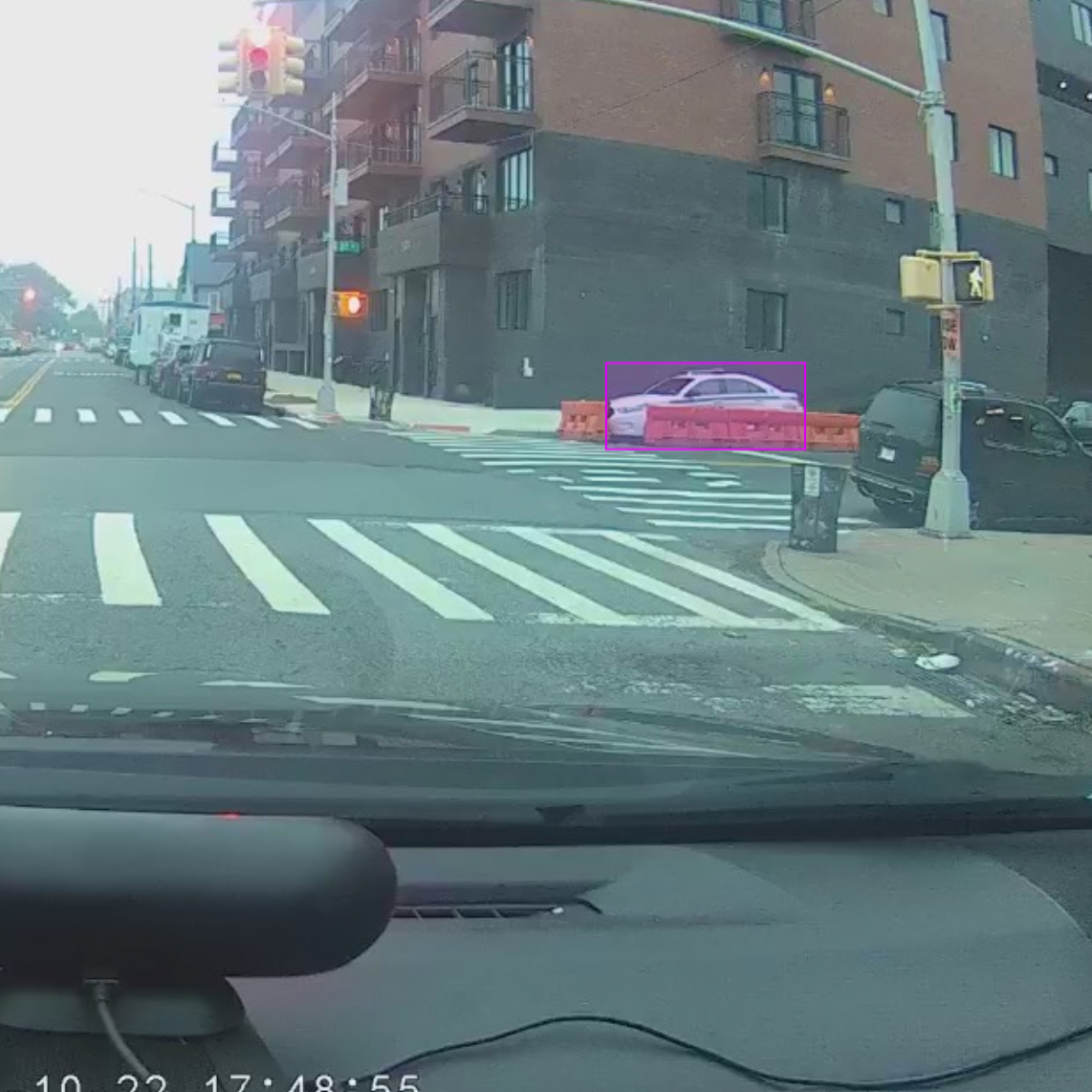} & 
            \includegraphics[width=.5\textwidth]{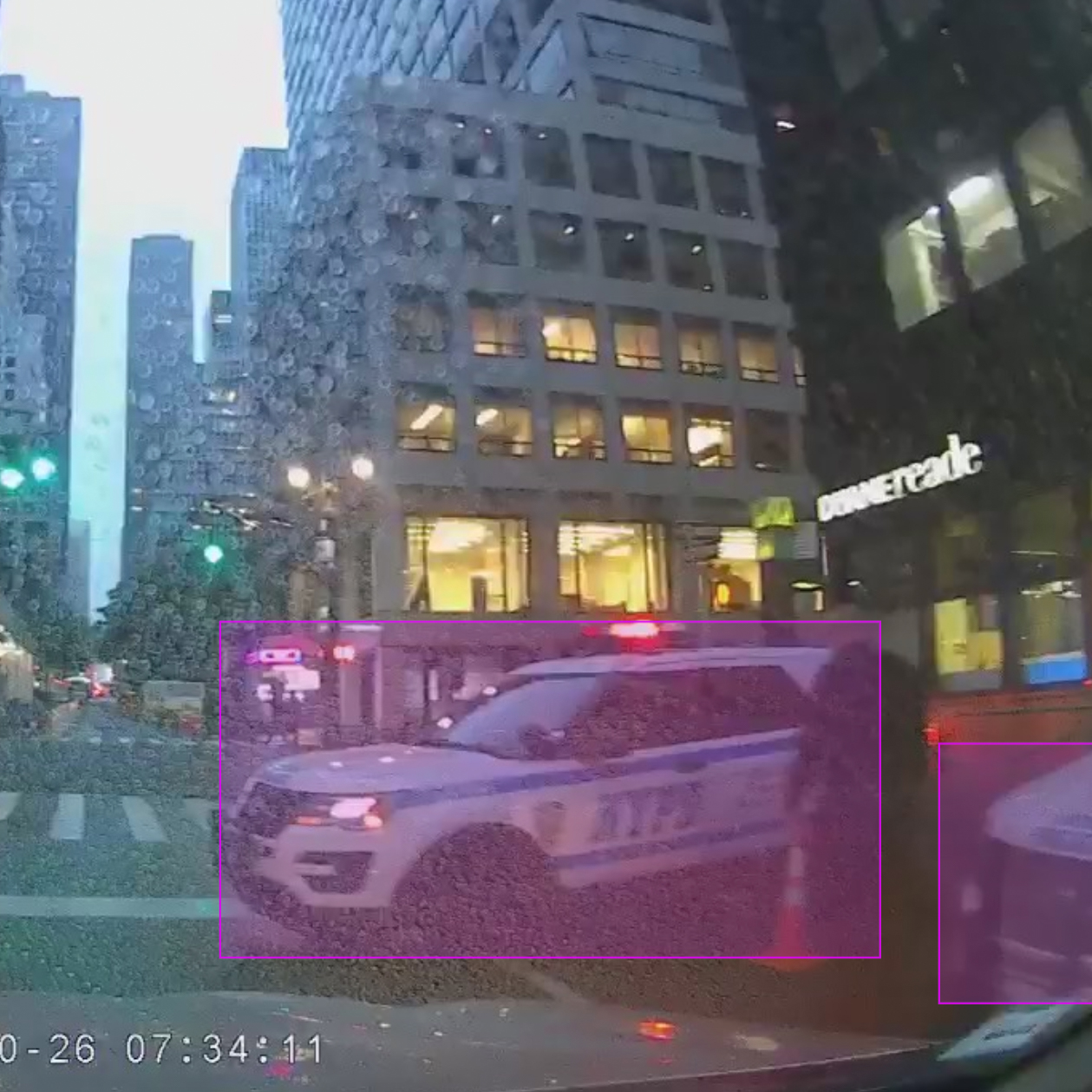} \\
            \includegraphics[width=.5\textwidth]{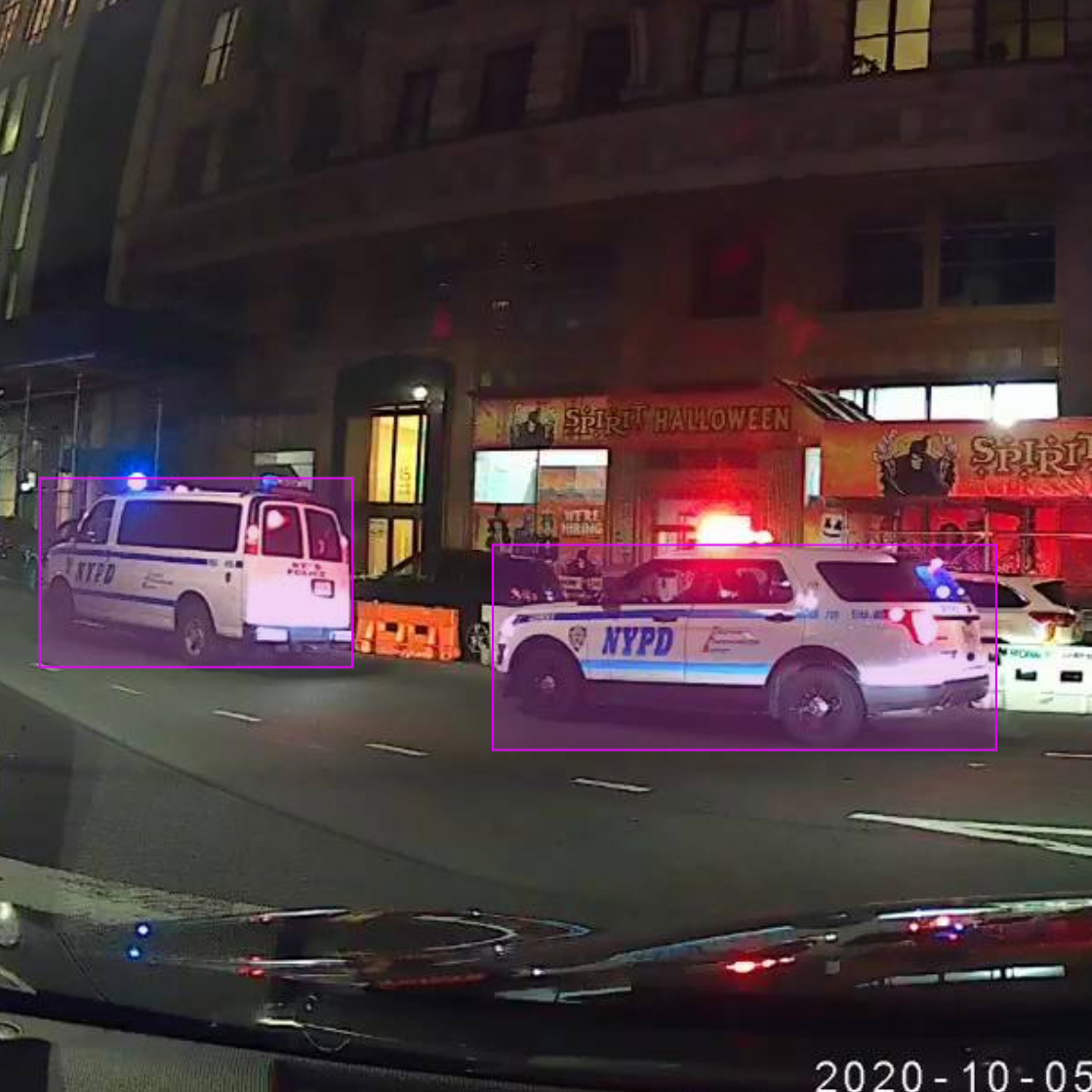} & 
            \includegraphics[width=.5\textwidth]{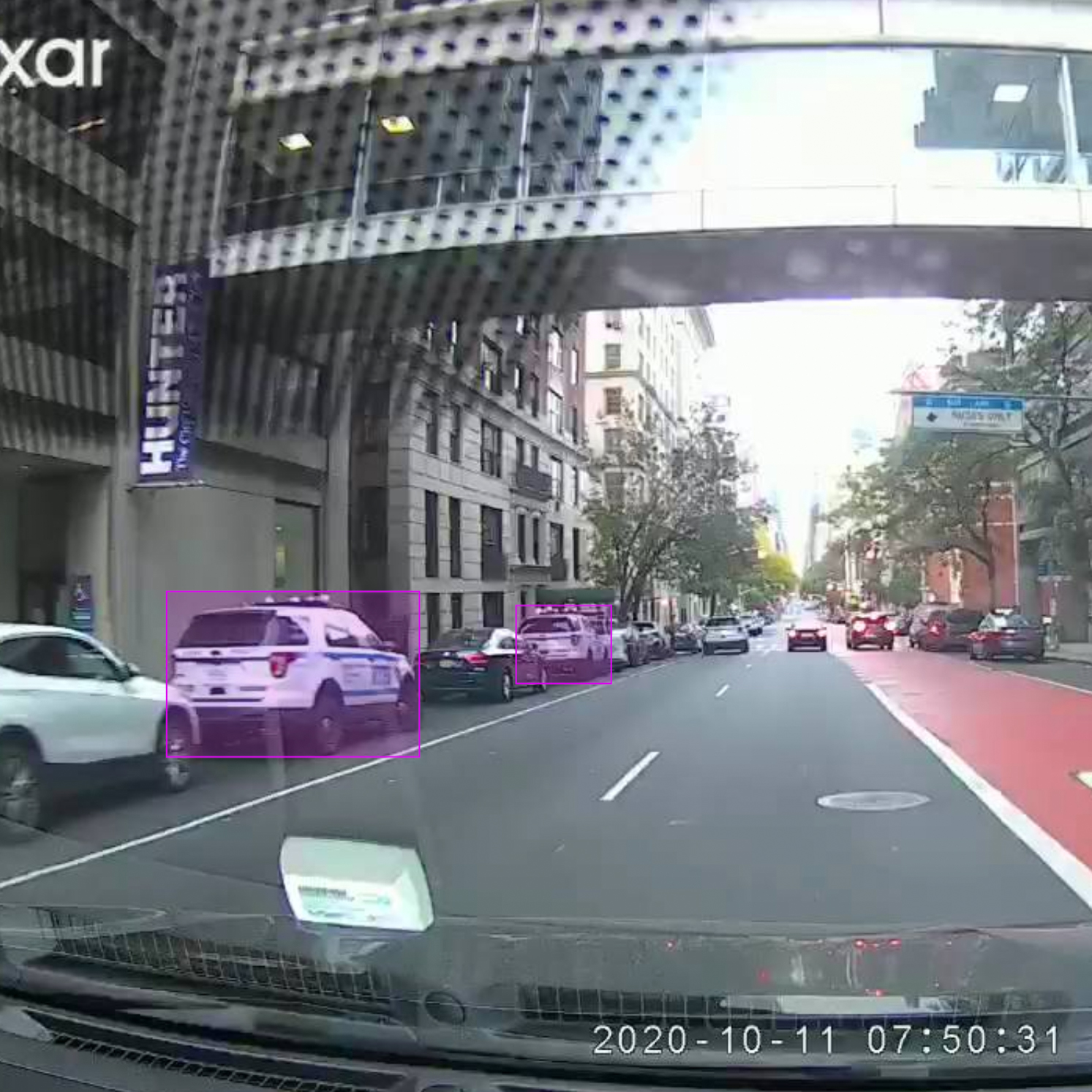} \\
         \end{tabular}
         \caption{True Positives}
         \label{fig:tp}
     \end{subfigure}
    \hfill
    \begin{subfigure}[b]{0.35\textwidth}
         \centering
         \begin{tabular}{ll}
            \includegraphics[width=.5\textwidth]{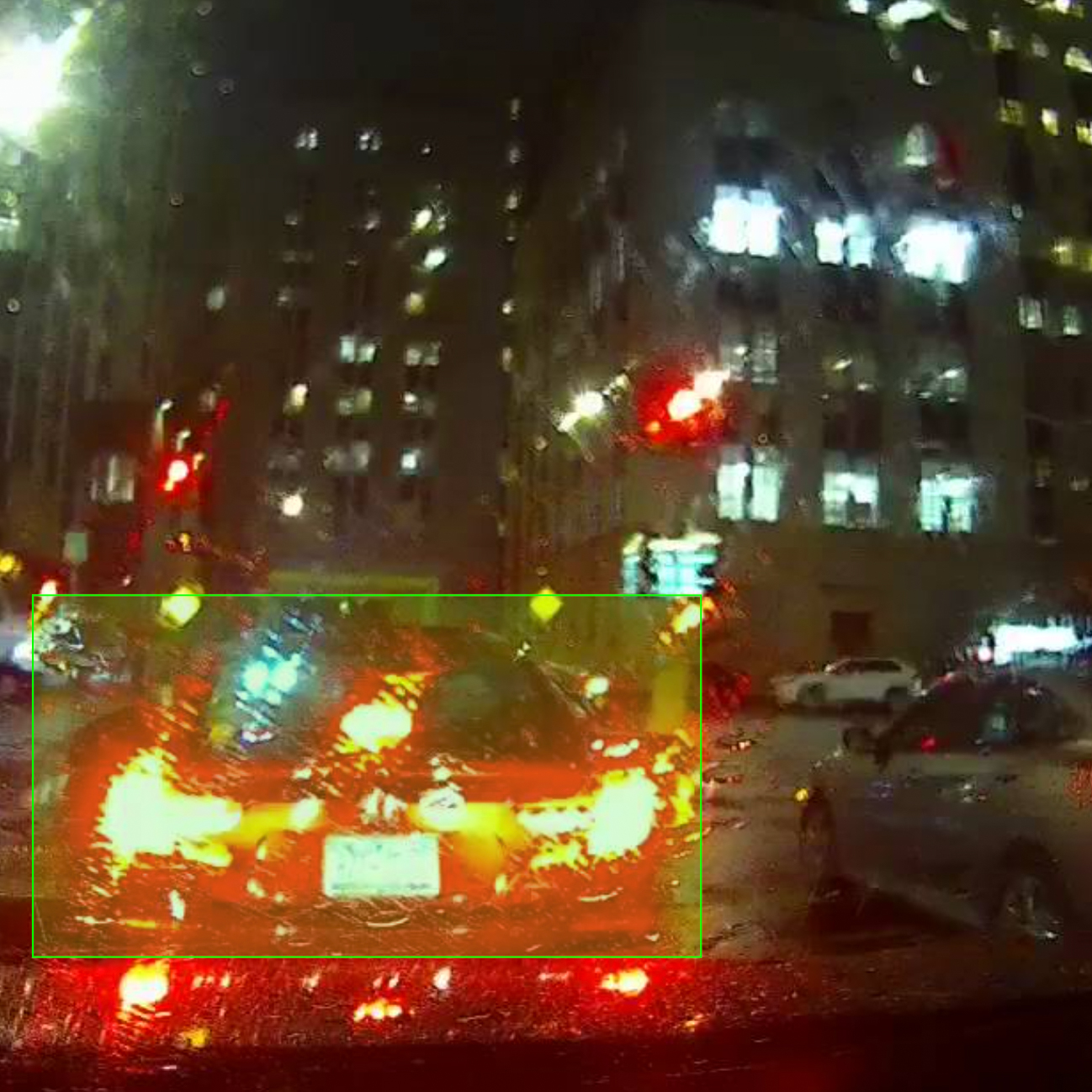} & 
            \includegraphics[width=.5\textwidth]{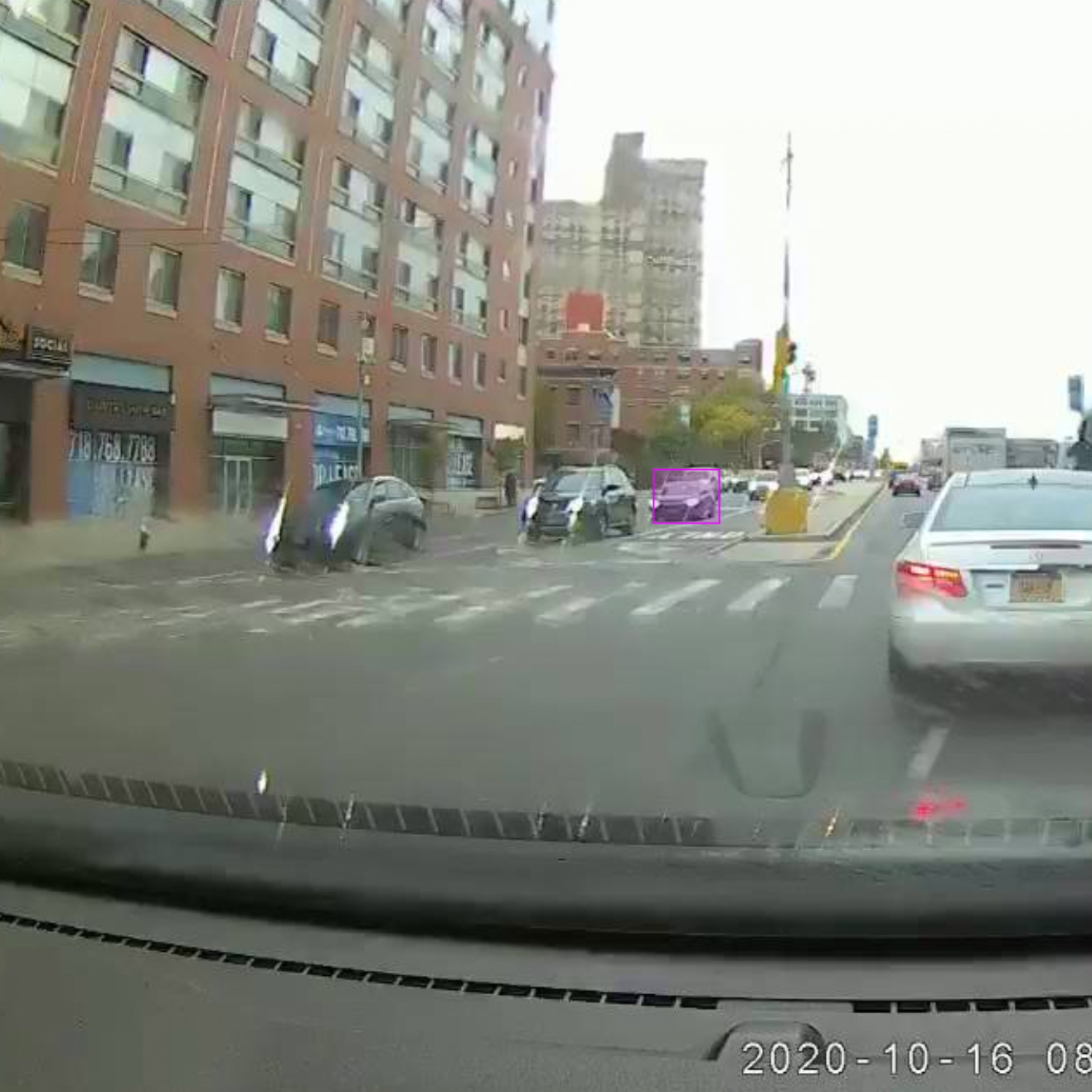} \\
            \includegraphics[width=.5\textwidth]{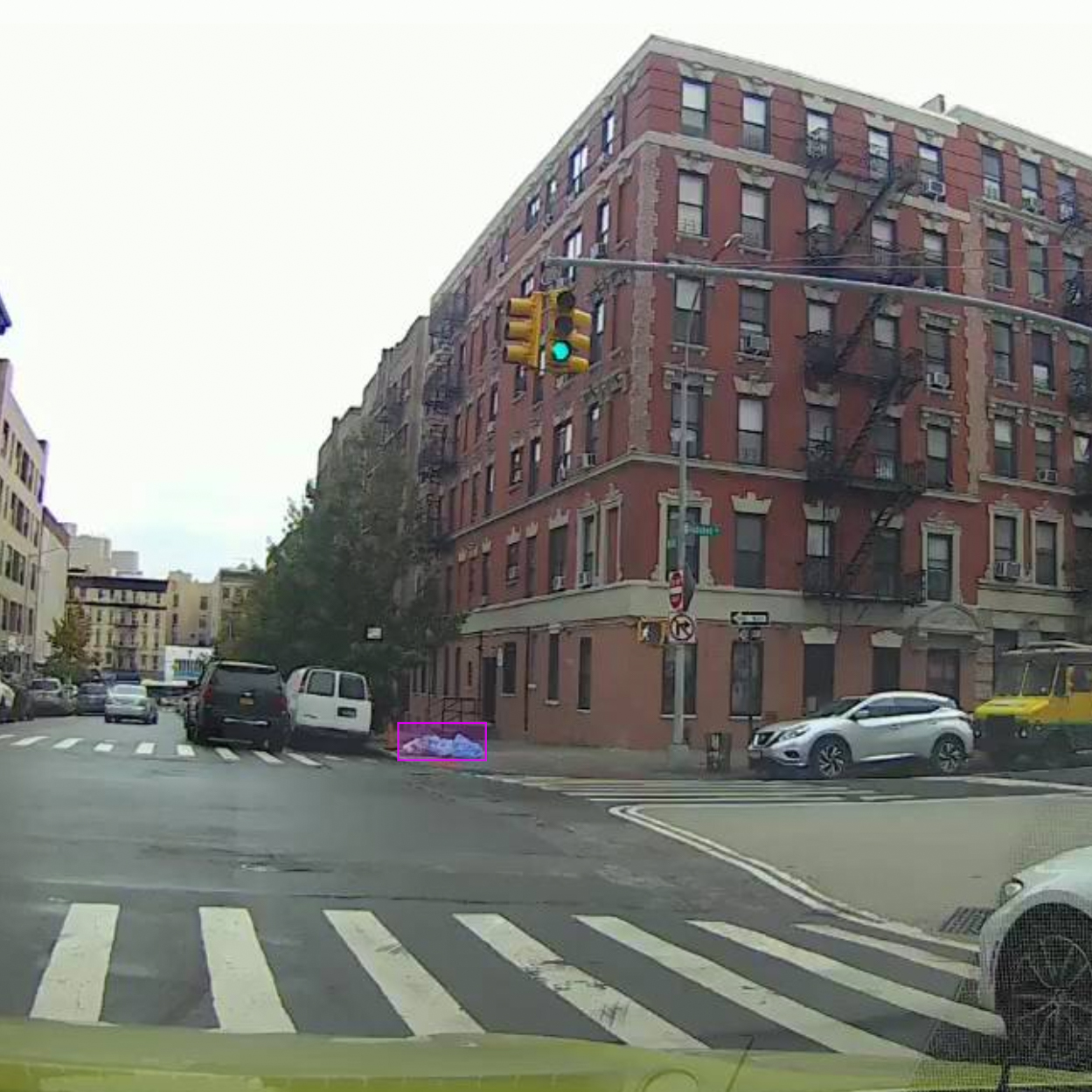} & 
            \includegraphics[width=.5\textwidth]{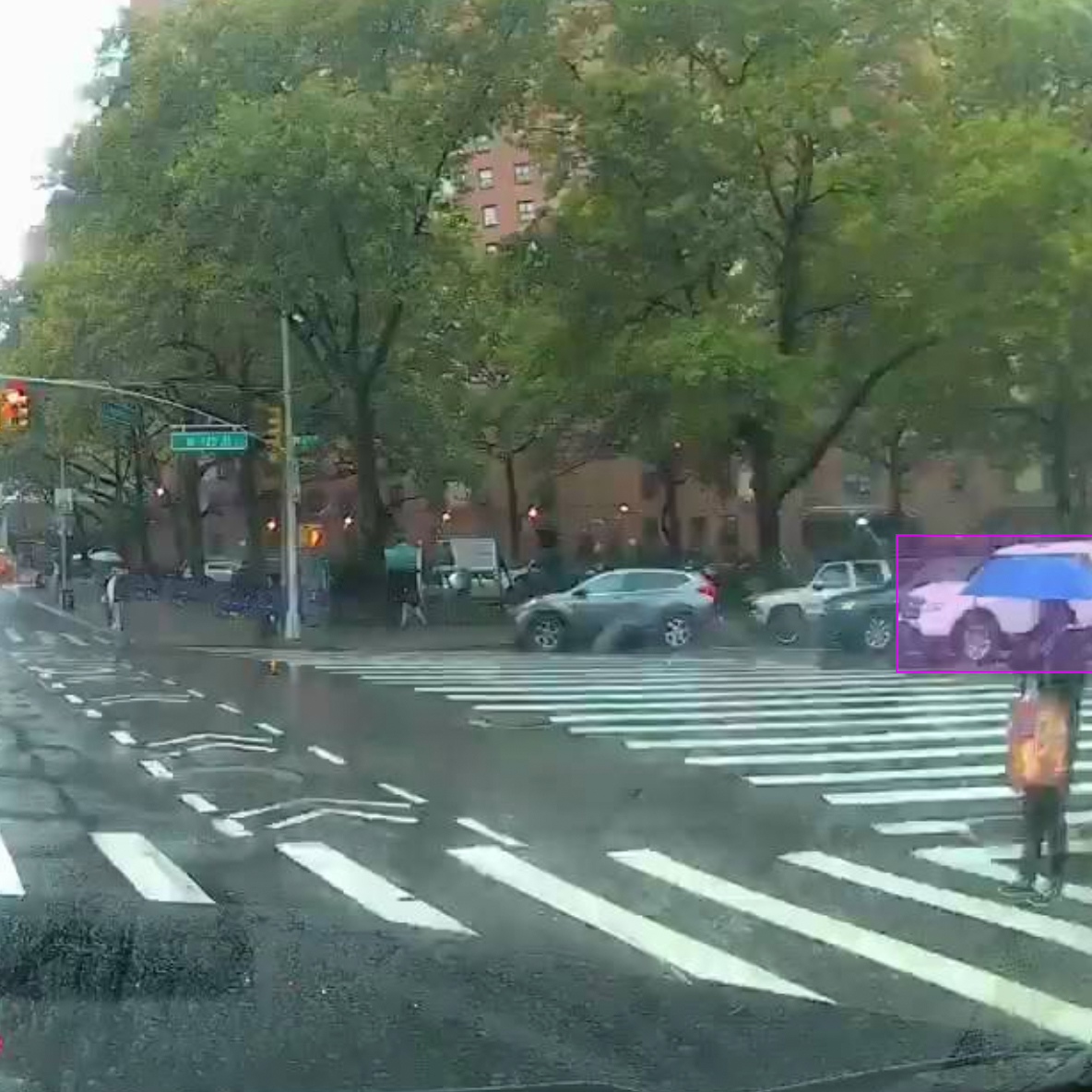} \\
         \end{tabular}
         \caption{False Positives}
         \label{fig:fp}
     \end{subfigure}
     \hfill
     \begin{subfigure}[b]{0.175\textwidth}
         \centering
         \begin{tabular}{l}
            \includegraphics[width=\textwidth]{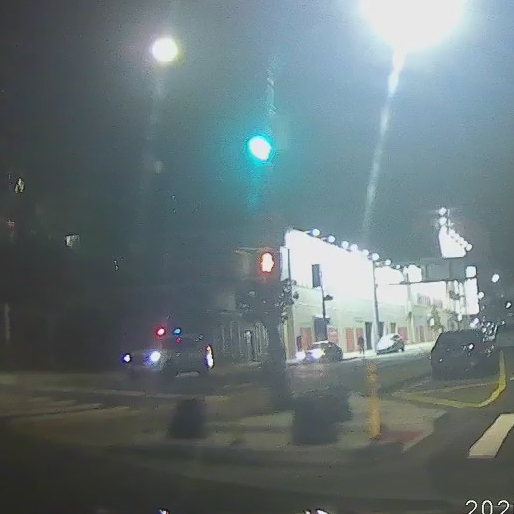} \\
            \includegraphics[width=\textwidth]{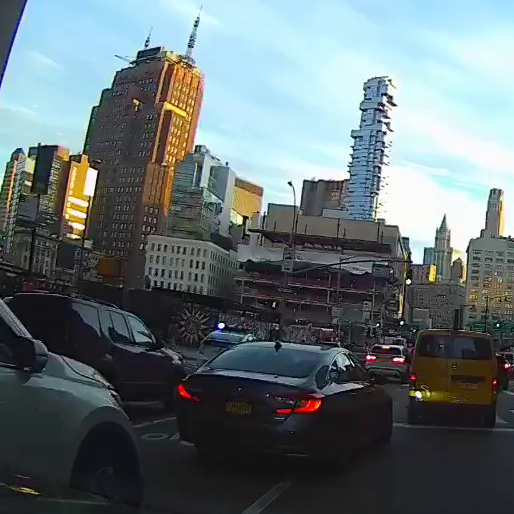} \\
         \end{tabular}
         \caption{False Negatives}
         \label{fig:fn}
     \end{subfigure}
    
    \caption{Examples of police vehicle detections. Notice the different times of day, differing camera perspectives, and multiple types of vehicles. Initially, the object detection model was easily confused by NYC taxis, buses, white sedans, and white SUVs. As more training data was added, the model was more difficult to confuse, with incorrect classifications occurring typically with optical illusions and poor light conditions. Note that images are cropped and zoomed to better depict annotated regions; raw images are 1280x720.} 
    \label{fig:detections}
    
\end{figure*}

Using the held-out test set, we first assess model performance both on the dataset as a whole, and then verify that the model achieves consistent performance across subsets of the dataset. Figure \ref{tab:test_set_performance_figure} shows the precision-recall curve and ROC curve for the test set, and Table S2 
reports average precision, AUC (i.e., AUROC), precision and recall. The model achieves high performance (test set AUC 0.99, average precision 0.82). This speaks to one benefit of our methodology: because we have chosen a relatively feasible object detection task (likelier easier than identifying surveillance cameras, as attempted in previous work~\cite{sheng_surveilling_2021}, which are much smaller) we are able to achieve high performance. Figure \ref{fig:detections} shows specific true positive, false positive, and false negative classifications on the dataset, demonstrating that the model can recognize police vehicles in a wide variety of images, but also that it can be confused by optical illusions and poor light conditions.

\setlength{\tabcolsep}{0.1em} %
\setlength{\tabcolsep}{0.5em} %

\begin{figure*}
    \begin{center}
        \includegraphics[width=\textwidth]{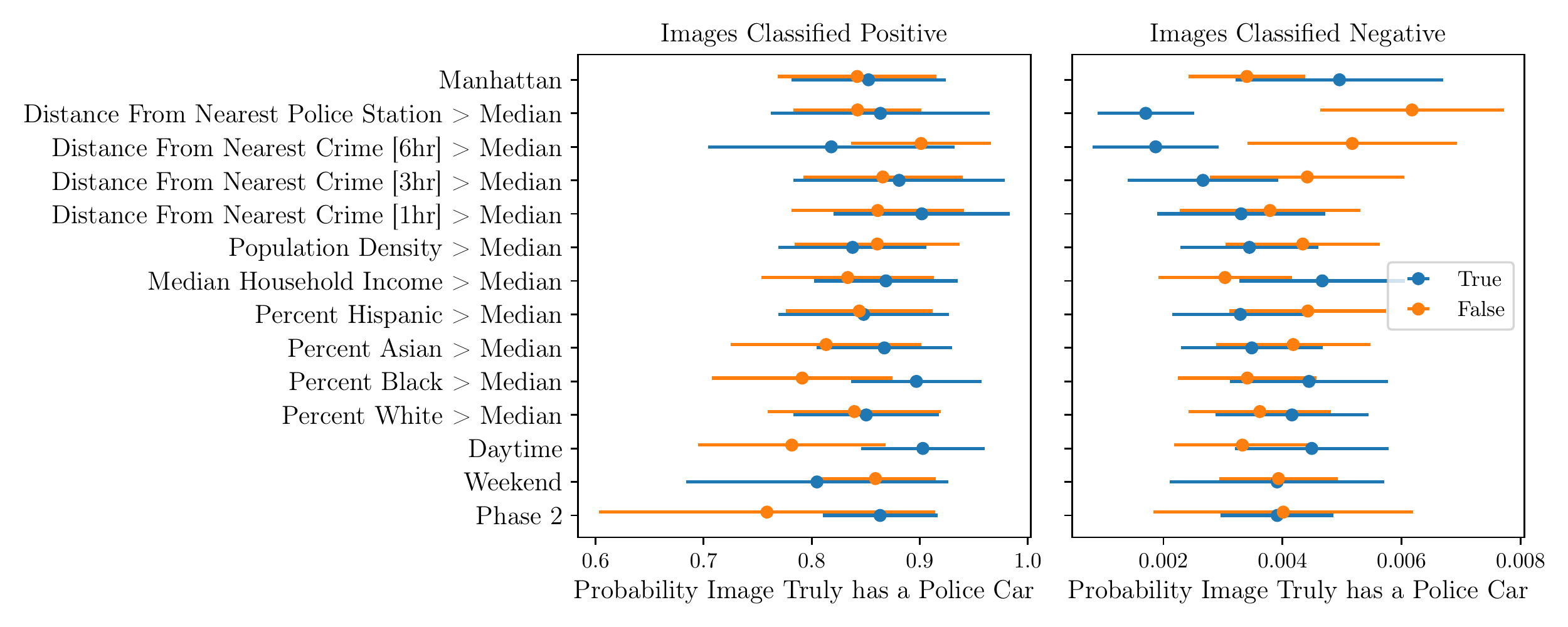}
        \caption{Calibration plots by subgroup, evaluated on the test set. The left plot shows $\text{Pr}(y=1|\hat y=1)$; the right subfigure shows $\text{Pr}(y=1|\hat y=0)$. Lines indicate uncertainty in estimates, calculated as 1.96 times the Bernoulli standard error; if the difference between the orange and blue dots is smaller than the uncertainty, it indicates that there is no statistically significant difference in model performance across the stratification.}
        \label{fig:calplots}
    \end{center}
\end{figure*}

Our estimation framework (\S \ref{sec:conceptual_framework}) requires that our model is \emph{calibrated} across subgroups --- that is, $\text{Pr}(y=1|\hat y=1)$ and $\text{Pr}(y=1|\hat y=0)$ remain similar across subgroups. We therefore assess whether this is the case using the test set, and as an additional validation assess model average precision and AUC across subgroups. Figure \ref{fig:calplots} shows that the model is calibrated across demographic variables (e.g. \% White, \% Black, population density, median household income, and Manhattan vs. non-Manhattan). The model is also calibrated across a number of other variables (whether the image is taken during the day, on the weekend, or during phase 1). Table S1 
shows that AUC and average precision remain consistently high across subgroups (AUC 0.98-0.99 for all subgroups; average precision 0.77-0.84). The only evidence of miscalibration we see does not occur in the variables over which we assess disparities, and thus does not threaten the validity of those estimates. Specifically, the model is somewhat miscalibrated by distance from nearest police stations: $\text{Pr}(y=1|\hat y=0)$ is higher for images closer to police stations. This is unsurprising, given the overdensity of police vehicles near police stations, and may also result from the distribution shift from training set to validation/testing set. We experiment with two fixes for this issue: first, we try upsampling images which the model misclassifies, but find it harms overall performance significantly. Second, we recalibrate the final model using distance from nearest police station as a covariate. This recalibration does not significantly affect our main results, as expected, so we do not use the recalibrated model for simplicity. There is also slight evidence of miscalibration by distance from nearest crime, but it is inconsistent and depends on the temporal threshold used for determining the nearest crime. 

Overall, the model is generally well-calibrated and achieves high performance across subgroups over which we assess disparities, as our mathematical framework requires. At our positive classification threshold of 0.77, the model classifies \numImagesAboveThreshold images (out of \numTotalImages  total) as containing a police vehicle.  From the validation set, we estimate $\widehat{\text{Pr}}(y=1|\hat y=1)$ and $\widehat{\text{Pr}}(y=1|\hat y=0)$, and use them as described in \S \ref{sec:conceptual_framework} to compute disparities in police deployments across demographic groups. 

As an additional validation, we assess whether the police detections correlate as expected with external data. Specifically, we assess whether Census Block Groups whose images have a lower mean distance to crime or to police stations are likelier to have police vehicles detected, as one would expect. We find that both correlations hold ($p<0.001$, Pearson correlation across Census Block Groups). This provides an additional validation beyond assessing model performance, since it suggests that police vehicle presence shows the same correlations with crime and police stations which we would expect police activity as a whole to display.

\section{Analysis of disparities}

\setlength{\tabcolsep}{0.1em} %
\begin{figure*}
    \centering
    \begin{subfigure}[b]{0.33\textwidth}
         \centering
         \includegraphics[width=\textwidth]{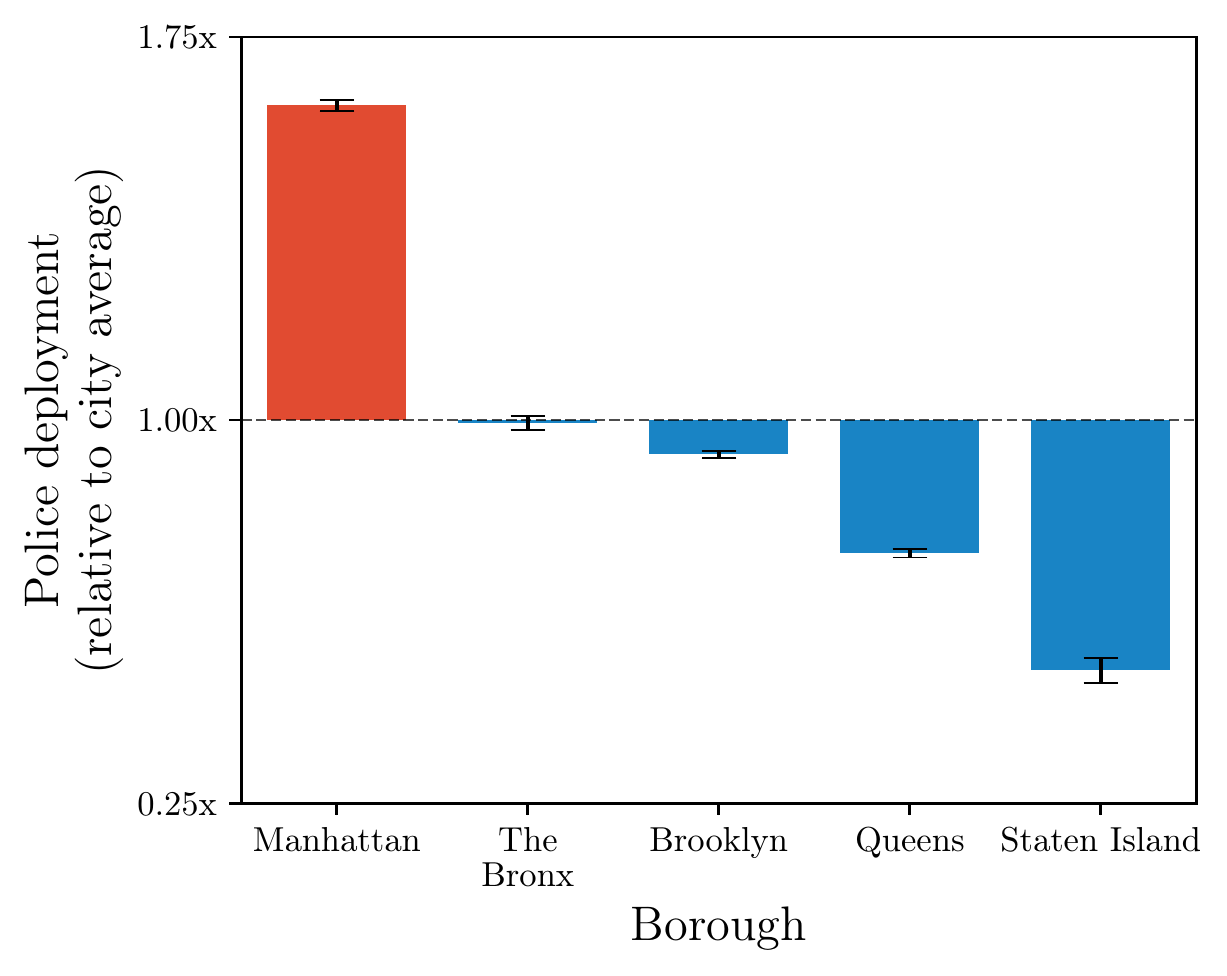}
         \caption{Deployments by borough.}
         \label{fig:zoning}
     \end{subfigure}
    \hfill
    \begin{subfigure}[b]{0.33\textwidth}
         \centering
         \includegraphics[width=\textwidth]{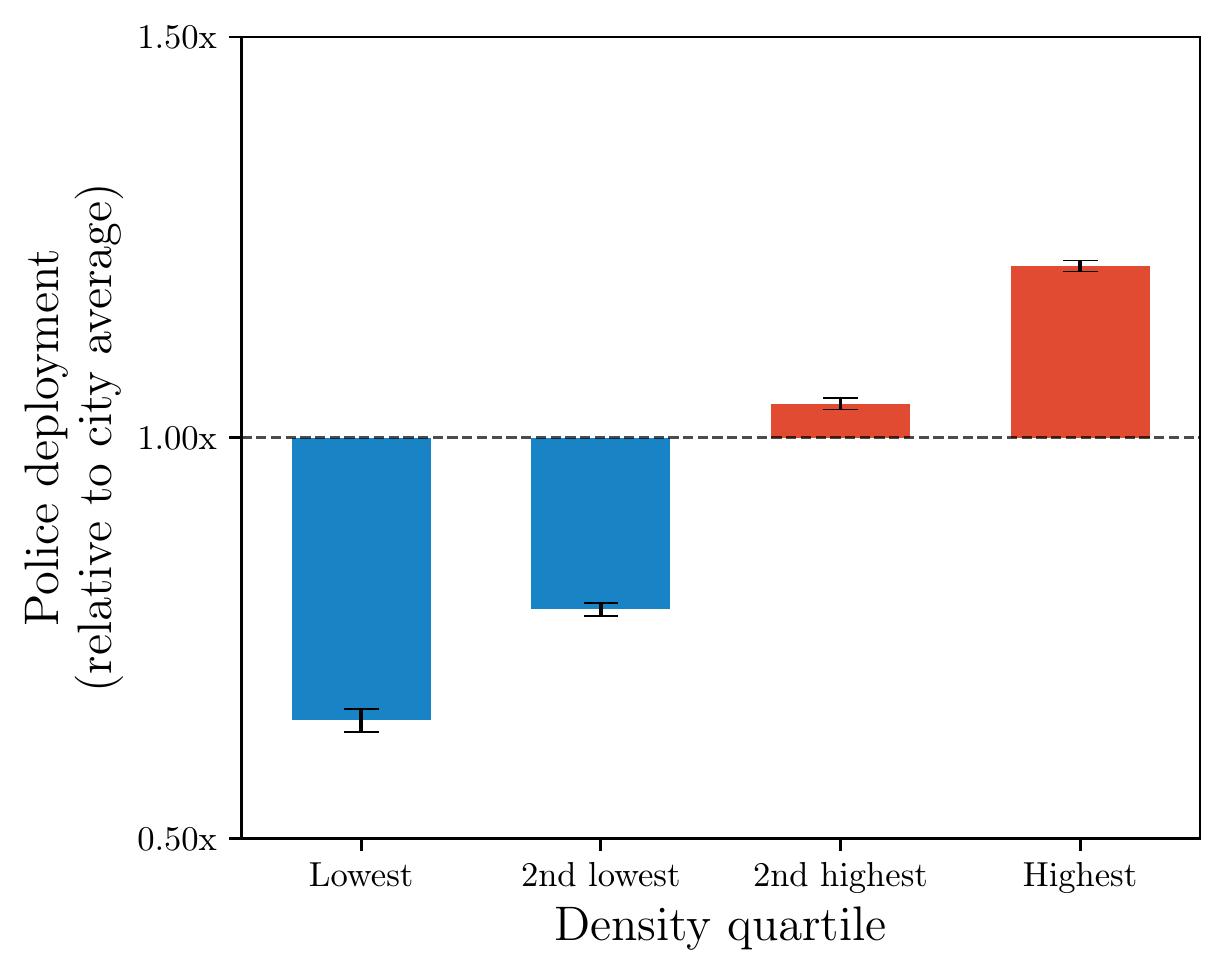}
         \caption{Deployments by population density.}
         \label{fig:density}
     \end{subfigure}
     \begin{subfigure}[b]{0.33\textwidth}
         \centering
         \includegraphics[width=\textwidth]{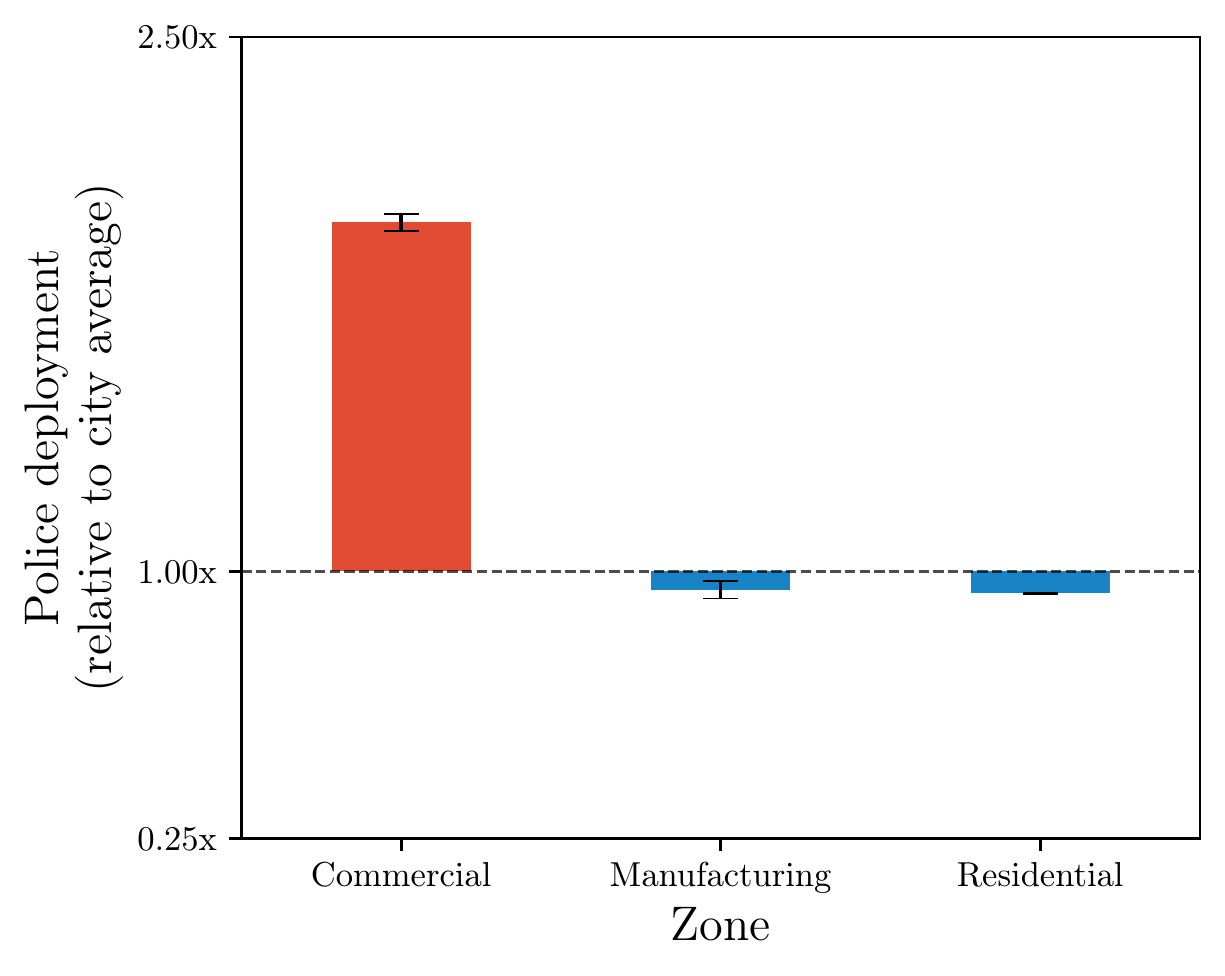}
         \caption{Deployments by zone type.}
         \label{fig:density-rzones}
     \end{subfigure}
    \caption{Disparities in police deployments by borough, density, and zone.} 
    \label{fig:borough-zone-barplots}
\end{figure*}
\setlength{\tabcolsep}{0.5em} %

We examine disparities in police deployments --- i.e., $\text{Pr}(y=1|A=a)$, computed as described in \S \ref{sec:conceptual_framework} --- by racial group (white, Black, Hispanic, and Asian), zone type (commercial, residential, and manufacturing), population density, median household income, borough, and neighborhood. Throughout the results, we express police deployments relative to the population-weighted city overall average: e.g., $1.6\times$ for a group indicates that deployment levels are 1.6 times the overall average. We compute error bars for all estimates by bootstrapping; reported errors are 1.96 times the standard deviation across bootstrapped datasets. 

\begin{figure}[h!]
    \begin{center}
  \includegraphics[width=0.5\textwidth]{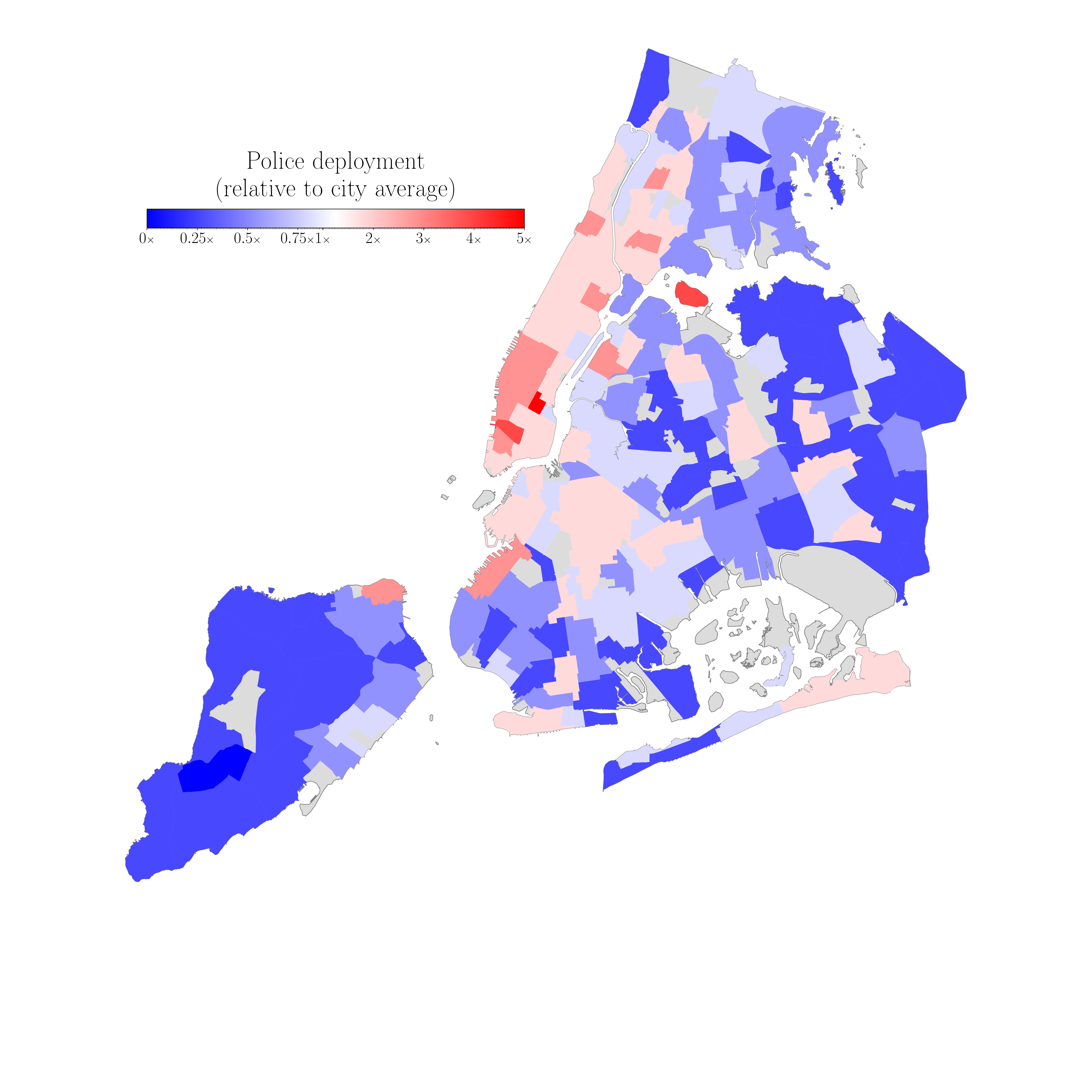}
  \caption{Map of police deployment throughout NYC, expressed relative to the city average. Grey areas are those with zero population in Census data, including airports, cemeteries, and parks.}
    \label{fig:pr-police-by-nta}
  \end{center}
\end{figure}

We find significant disparities in police deployments across boroughs (Figure \ref{fig:borough-zone-barplots}) and neighborhoods (Figure \ref{fig:pr-police-by-nta}). The borough with the highest police deployments (Manhattan) has levels $1.62\pm0.01\times$ the city average; the borough with the lowest police deployments (Staten Island) has levels $0.51\pm0.02\times$ the city average, a more than 3-fold difference. Similarly, the neighborhood with the highest police deployment levels (Gramercy) has levels $4.28\pm0.10\times$ the city average; the neighborhood with the lowest police deployments (Arden Heights-Rossville) has levels $0.23\pm 0.02\times$ the city average, a nearly 20-fold difference. Examining the neighborhoods with the highest police levels reveals a striking mixture of two types of neighborhoods: some of the wealthiest areas in New York (Hudson Yards, once described as a ``billionaire's fantasy city''~\cite{hudson_yards}) and some of the most heavily policed (Rikers Island, which houses a prison complex).

High police deployment levels in some neighborhoods may be driven by idiosyncratic neighborhood attributes --- for example, Gramercy is home to a major police training center on two busy thoroughfares. To more systematically examine spatial disparities, therefore, we examine correlations with zone and density (Figure \ref{fig:borough-zone-barplots}). Commercial zones have much higher police deployments ($1.98\pm0.03\times$ the city average) than residential or manufacturing zones. Police deployments also increase dramatically with population density, from $0.65\pm0.01\times$ the city average in areas in the lowest quartile of density to $1.21\pm0.01\times$ the city average in areas in the highest quartile of density.\footnote{We compute quartile boundaries at the image level.} Overall, this analysis suggests that one place high police levels are observed is dense, commercial regions of downtown Manhattan. %

\begin{figure*}
    \centering
    \begin{subfigure}[b]{0.48\textwidth}
         \centering
         \includegraphics[width=\textwidth]{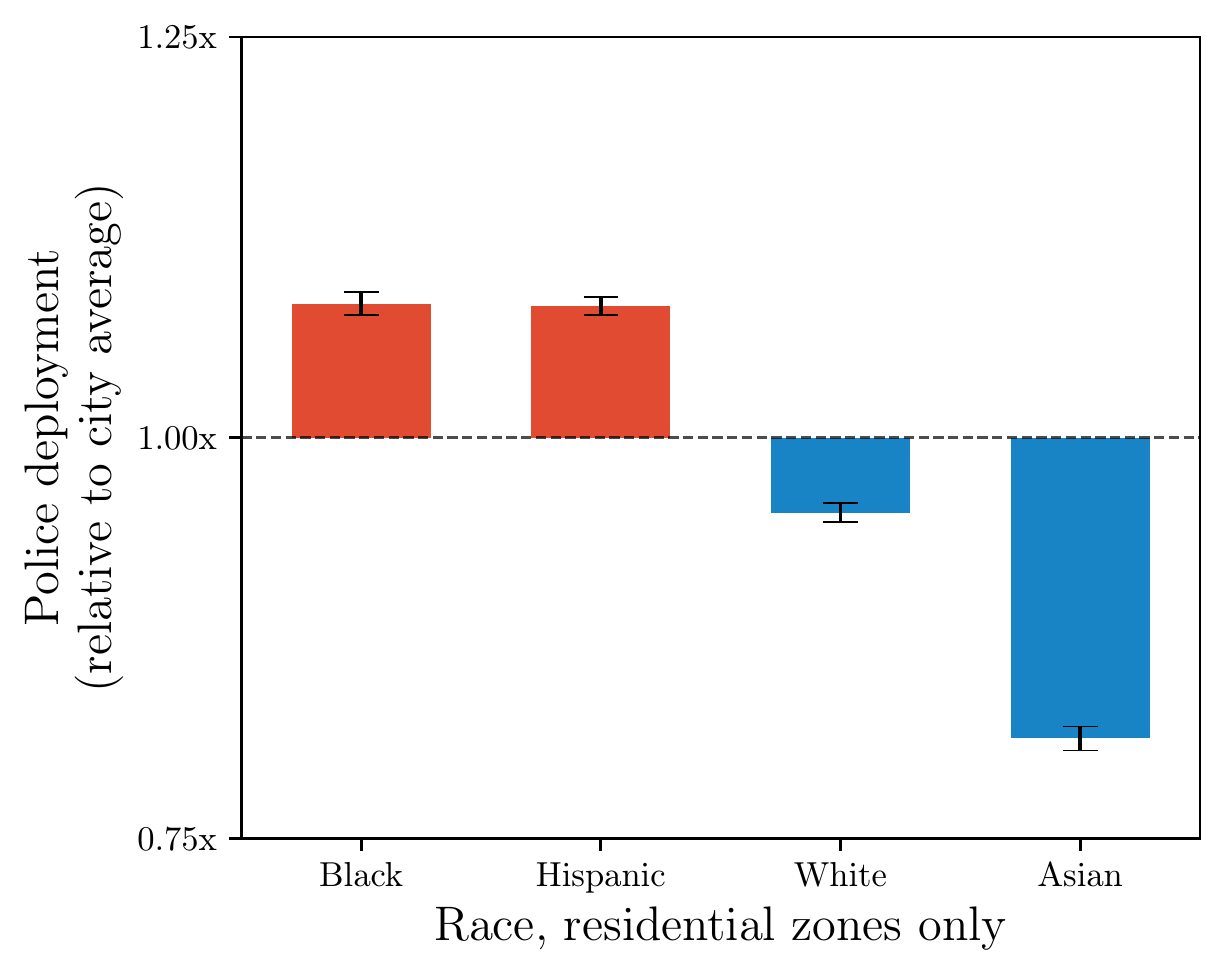}
         \caption{Deployments by race group.}
         \label{fig:race}
     \end{subfigure}
    \hfill
    \begin{subfigure}[b]{0.48\textwidth}
         \centering
         \includegraphics[width=\textwidth]{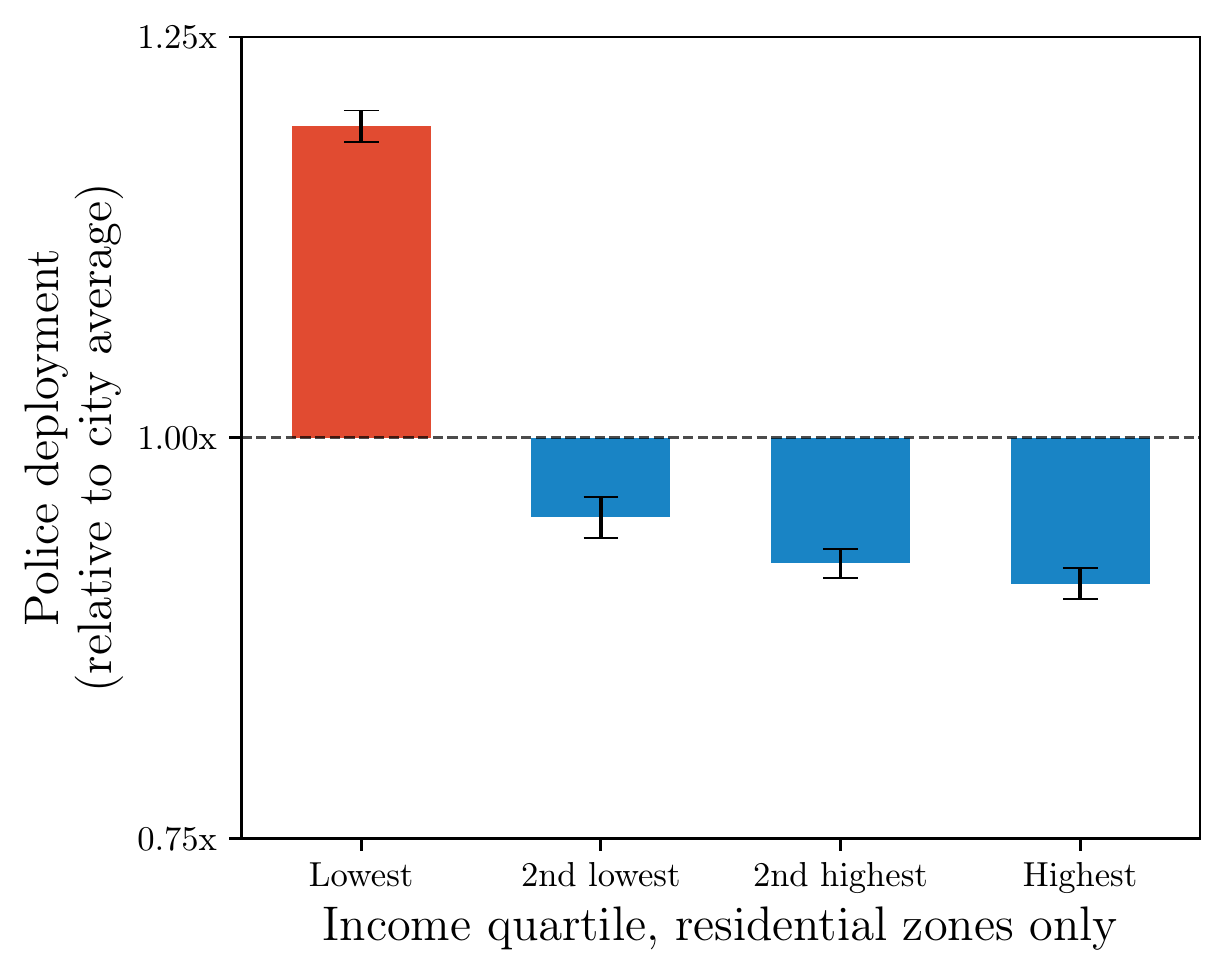}
         \caption{Deployments by income quartile.}
         \label{fig:income}
     \end{subfigure}
    \caption{Disparities in police deployments by race group (a) and median household income (b). Plots are made using data from Census Block Groups in residential zones. Figure S2 %
    shows the corresponding plots without filtering for residential zones; results for race are qualitatively similar, but results for income show a U-shaped trend, for reasons we discuss in the main text.} 
    \label{fig:demographics-barplots}
\end{figure*}

We next analyze disparities in police deployments by race and socioeconomic status (Figure \ref{fig:demographics-barplots}a), restricting the analysis to residential zones to capture where people actually live. We observe substantial racial disparities. Black and Hispanic residents face police deployment levels $1.08\pm0.01\times$ the city average, as compared to $0.95\pm0.004\times$ for white residents and $0.81\pm0.01\times$ for Asian residents, a 34\% discrepancy between the highest and lowest race groups. This is consistent with previously observed disparities in policing in New York City~\cite{gelman2007analysis,pierson2018fast,jung2018omitted} where Black and Hispanic residents were more heavily policed. (In Figure S2 
we plot racial disparities without restricting to residential zones, yielding consistent results --- Black and Hispanic residents face higher police deployments than white and Hispanic residents --- although the disparities narrow somewhat, to a 19\% gap between the highest and lowest race groups.)

We also observe disparities by Census Block Group (CBG) median income (Figure \ref{fig:demographics-barplots}b). Again restricting to residential zones, police deployments increase as income decreases: residents of CBGs in the lowest median household income quartile face police deployments which are $1.19\pm0.01\times$ the city average, while residents of CBGs in the highest median household income quartile face police deployments only $0.91\pm0.01\times$ the city average. Interestingly, the relationship between income and deployment becomes U-shaped if we do not restrict to residential zones (Figure S2 
): neighborhoods in the \emph{top} quartile of median income, and the \emph{bottom} quartile of median income, both experience higher police deployments than neighborhoods in the middle two quartiles. In other words, top-quartile income areas go from having the lowest police deployments (when we analyze only residential zones) to the highest (when we analyze all areas). The difference likely occurs because of higher-income commercial zones which are heavily policed. %

We note that we report racial and socioeconomic disparities without attempting to control for other covariates for two reasons. First, if New York City residents of different races face different levels of police deployments, that disparate impact is itself important; it can also bias algorithms trained on downstream policing data irrespective of the true causal mechanism. Second, controlling for other factors in policing data can be difficult to interpret, introducing concerns about omitted variable bias and model misspecification which make it difficult to identify which factor is truly the ``cause'' of higher police deployments. For the sake of transparency and simplicity, therefore, we report results by stratifying each variable separately, noting that these disparities are themselves important but that multiple causal mechanisms may underlie them. %

\setlength{\tabcolsep}{0.1em} %
\setlength{\tabcolsep}{0.5em} %

\section{Discussion}

We present a novel methodology for studying disparities in police deployments. We show that a deep learning model can be applied to tens of millions of dashcam images to identify police vehicles with high accuracy, and introduce a principled framework for estimating disparities in a way that mitigates data and model biases. We find substantial spatial inequality in where police vehicles are deployed. Residential areas with higher proportions of Black and Hispanic residents, or lower-income residents, have significantly higher police deployments. But so, too, do dense, wealthy commercial areas in downtown Manhattan like Hudson Yards; this finding that police deployments can be high in wealthy neighborhoods is consistent with past work showing that gentrification correlates with increased police presence~\cite{beck2020policing}. Overall, our analysis speaks to a complex, multifaceted picture of police deployment in New York.

We employ a number of techniques to reduce bias in our pipeline, providing a template for future computational social science research on non-representative dashcam data. However, two unavoidable limitations are important to acknowledge. First, while we reweight the sample such that each Census area is weighted proportional to its population, it is possible that the sample of images we have \emph{within each Census area} is biased. In other words, the police vehicle frequency within our Nexar data sample for each Census area may not accurately represent the true police vehicle frequency within that Census area. News articles document, for example, how protests after George Floyd's death blocked traffic throughout New York City~\cite{nytimes_floyd_protests}, which would obviously preclude getting dashcam images from some areas; it is also plausible that ride-share drivers would seek to avoid areas with protest activity. Since protest activity likely correlates both with our data availability and with police vehicle presence, this could plausibly bias our estimates. We mitigate the specific concern about protests by conducting our analysis of disparities using data only from October 2020 onward, when protest activity was reduced. Nonetheless, the sampling process for the Nexar dataset is somewhat opaque and the time period analyzed was full of dramatic changes in social distancing policy and responses to policing so it would be unwise to assume that no other biases remain. Similarly, it is possible the time period sampled captures trends specific to the COVID-19 pandemic. At the same time, that time period itself is of interest because it is known that there were racial disparities in policing through the pandemic.~\citet{kajeepeta_policing_2022} find that from March to May 2020, a one standard deviation increase in percentage of Black residents was associated with a 73\% increase in the COVID-19-specific summons rate and a 34\% increase in the public health and nuisance arrest rate. Finally, we assume that people in the same Census area have the same probability of seeing a police vehicle irrespective of their group; while the Census areas we analyze (Census Block Groups) are quite small, rendering this assumption more plausible, it is possible there is additional variation correlated with group within Census areas. %

A second important limitation is that identifiable police vehicles are only a partial proxy for all policing activity. They do not capture, for example, officers on foot or unmarked vehicles. We did attempt to train a computer vision model to identify officers on foot; however, we found that the annotations we obtained were considerably lower quality than annotations for police vehicles, with annotators frequently confusing police officers with construction workers, cyclists with reflective vests, and other individuals in dark clothing. Given this, we opted to focus our analysis on police vehicles to ensure we could train a classifier with high accuracy. The high performance of the classifier likely stems in part from the iconic and unique branding of NYPD vehicles, which makes them very easy to spot and label in training images. While it is important to be aware that our analysis captures only one dimension of police activity, police vehicles are, themselves, an important indicator of policing activity: for example, they carry out traffic stops, which are one of the most common ways that Americans interact with police~\cite{pierson2020large}.

We assert our approach and results constitute a key step towards transparency in policing, proving it is possible to derive deployment patterns from dashcam data. Our results have several implications for equity and algorithmic fairness. First, our work reveals substantial inequality in police vehicle deployments. Police vehicles are more heavily deployed in neighborhoods with larger Black and Hispanic populations, recapitulating historic patterns of inequality in New York policing~\cite{goel2016precinct,pierson2018fast,gelman2007analysis}. These disparities are themselves of concern, and could also propagate into algorithms which made use of data from outcomes downstream of deployments, including arrests~\cite{laura2016public,corbett2017algorithmic,corbett2016computer,brookingspredictivepolicing,lum2016predict,baltimoresunpredictivepolicing,richardson2019dirty,selbst2017disparate}.%

Second, our work highlights the potential of dashcam data for auditing government agencies for efficiency and equity, including but not limited to the police, as well as for computational social science more broadly. For example, similar methodology could also be used to identify double parking issues,  dynamic obstacles to accessibility, or weather-induced hazards, and then to study disparities in government service response times, following previous work~\cite{lauferauditing}. Dashcam data is becoming increasingly available~\cite{Madhavan:EECS-2017-113,peng_visda_2018}, and dashcam is being postulated as a medium for future work~\cite{li_assessing_2015,li_examining_2021}. Our methods could easily be extended to other cities or detection tasks.  

Finally, our work serves as a call for the police themselves to release better deployment data. In the past police have resisted this~\cite{nytimes2017deployment} on the grounds that it would undermine public safety. But this claim stretches plausibility on multiple grounds: even \emph{aggregated} deployment data would be very useful for detecting disparities, as we show and past work has also found~\cite{nytimes_bronx_article,nytimes_bronx_article_2}. And far more granular data on many measures of policing activity is already publicly available --- including stops, searches, arrests, citations, and shootings~\cite{pierson2020large,washington_post_police_shootings} --- making it unclear how aggregated deployment data uniquely compromises police operations. The benefits would be substantial: public release of aggregated deployment data would democratize the ability to audit the police. This project was performed using highly specialized resources --- terabytes of data, a customized annotation pipeline and computer hardware, and hundreds of hours of model training --- and even then incurred unavoidable biases. This pipeline is not, in short, a realistic way for the average citizen to monitor the police. \emph{Quis custodiet ipsos custodes?} The answer cannot be ``only those with 25 million dashcam images.''%

\paragraph{Code availability:} Code used in this analysis is available at this GitHub repository: \href{https://github.com/mattwfranchi/police-deployment-patterns}{https://github.com/mattwfranchi/police-\newline deployment-patterns}. 

\paragraph{Dataset availability:}
It is our intent to share the data from this paper with other researchers, with protections to prevent abuse. To learn how to access to data from the study, see \href{https://github.com/mattwfranchi/police-deployment-patterns/wiki/Dataset-Access}{https://github.com/\newline mattwfranchi/police-deployment-patterns/wiki/Dataset-Access}.

\paragraph{Acknowledgments:} The authors thank Serina Chang, Nikhil Garg, Nakia Kenon, Hao Sheng, and Maria Teresa Parreira for helpful comments, and  Gabriel Agostini and Ilan Mandel for assistance with data and analysis. The Nexar dashcam dataset used in this study was licensed for use as part of NSF IIS-2028009 RAPID: Tracking Urban Mobility and Occupancy under Social Distancing Policy. WJ was supported by an Amazon Research Award. EP was supported by a Google Research Scholar award, an NSF CAREER award, a CIFAR Azrieli Global scholarship, a LinkedIn Research Award, and a Future Fund Regrant. JZ was partially supported by the United States Air Force and DARPA under contracts FA8750-20-C-0156, FA8750-20-C-0074, and FA8750-20-C0155 (SDCPS Program).

\balance
\bibliographystyle{ACM-Reference-Format}
\bibliography{Facct-Paper, matt_zotero_sources_copy}

\appendix 

\newpage
\section{Supplementary Figures}

\renewcommand\thefigure{S\arabic{figure}}
\renewcommand\thetable{S\arabic{table}}
\renewcommand{\figurename}{Figure}
\renewcommand{\tablename}{Table}
\setcounter{figure}{0}
\setcounter{table}{0}
\begin{figure*}[b]
\captionsetup{justification=centering}
    \begin{subfigure}[b]{.99\textwidth}
        \centering
        \includegraphics[trim={0 3.5cm 0 2.5cm},clip,width=\textwidth]{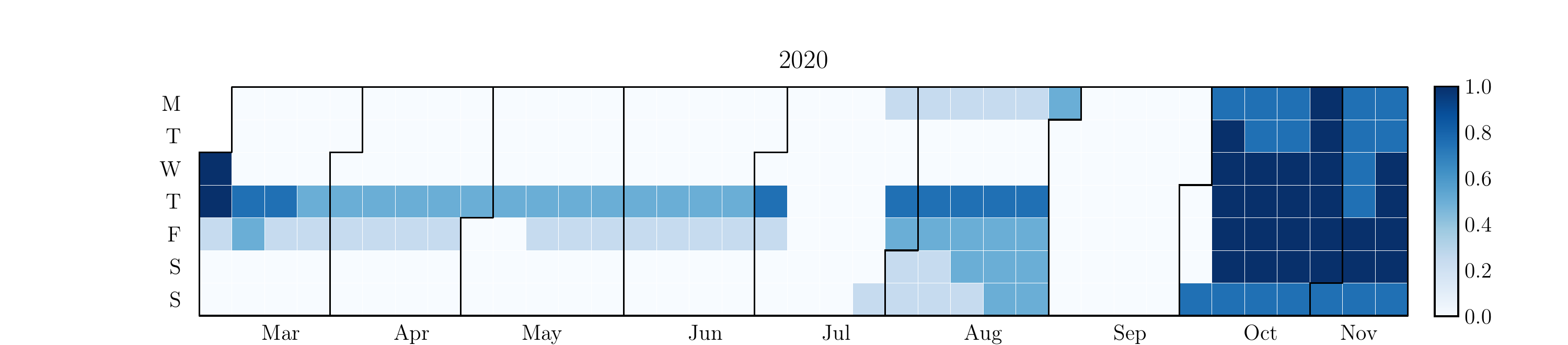}
        \caption{Calendar heatmap that shows data availability by day of month.\\ The heatmap is colored based on percentiles rather than raw values.}
        \label{fig:monthly}
    \end{subfigure} 

    \begin{subfigure}[b]{\textwidth}
        \centering
        \includegraphics[width=.99\textwidth]{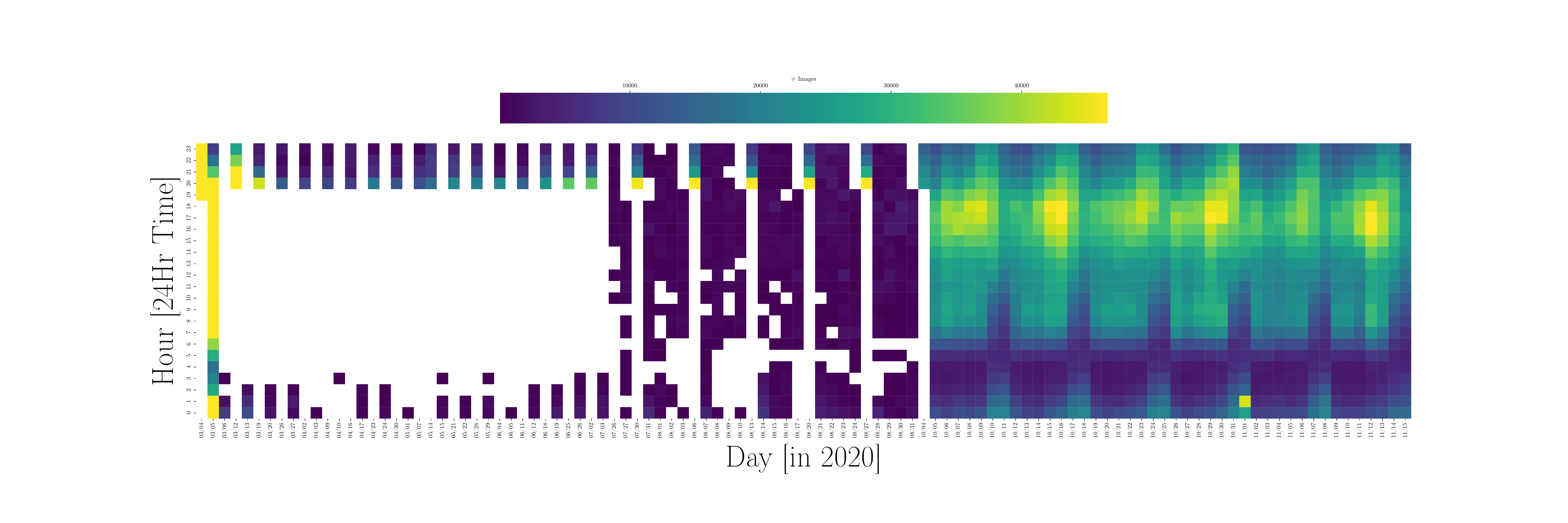}
        \caption{Calendar heatmap that shows sampling density by hour and day of coverage.}
        \label{fig:data-availability-calendar}
    \end{subfigure}
    \centering
    \caption{Temporal data availability.} 
    \label{fig:temporal-availability}
\end{figure*}

\setlength{\tabcolsep}{0.1em} %
\begin{figure*}[b]
    \centering
    \begin{subfigure}[t]{0.43\textwidth}
         \centering
         \includegraphics[width=\textwidth]{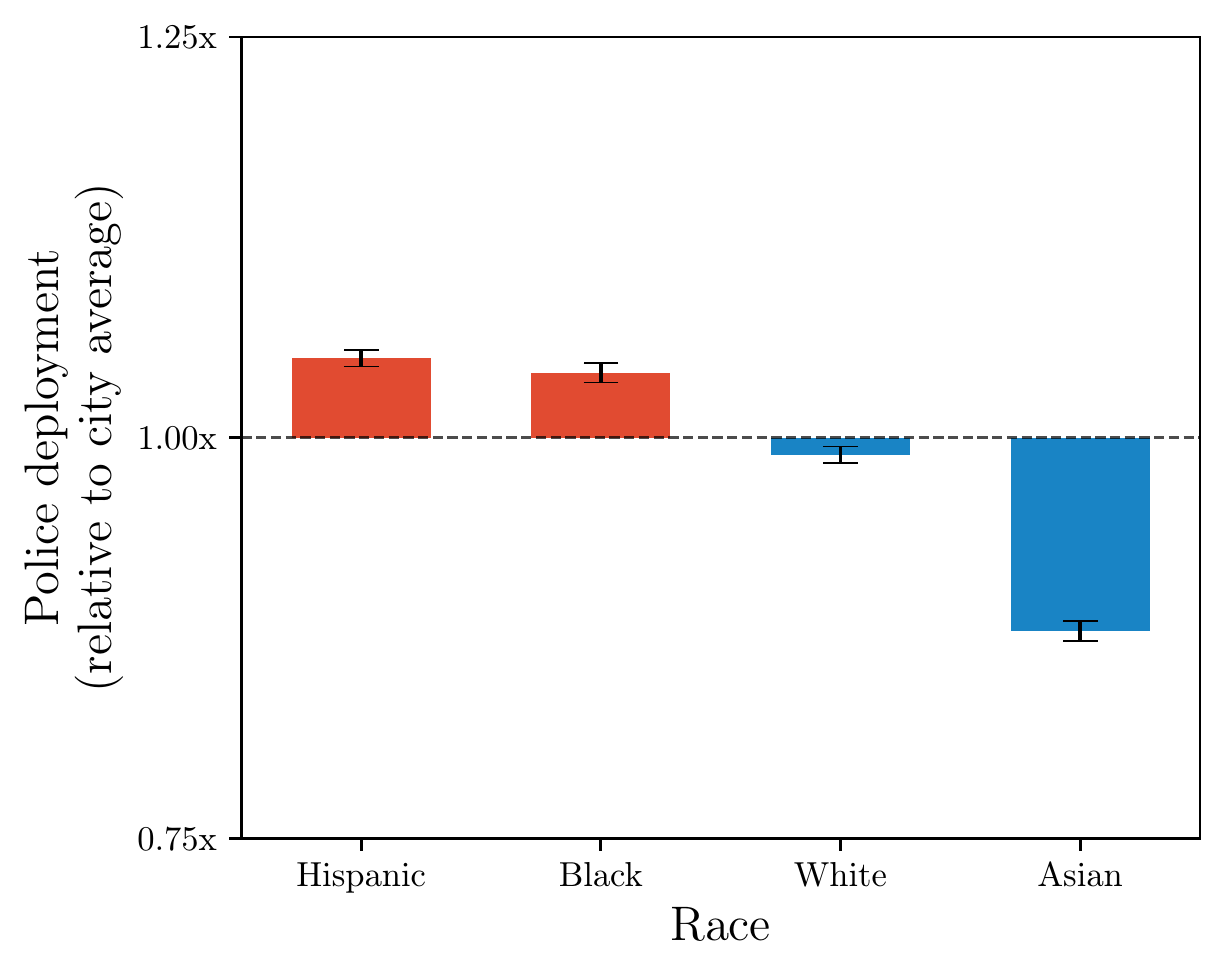}
         \caption{Deployments by race group.}
         \label{fig:race}
     \end{subfigure}
    \begin{subfigure}[t]{0.43\textwidth}
         \centering
\includegraphics[width=\textwidth]{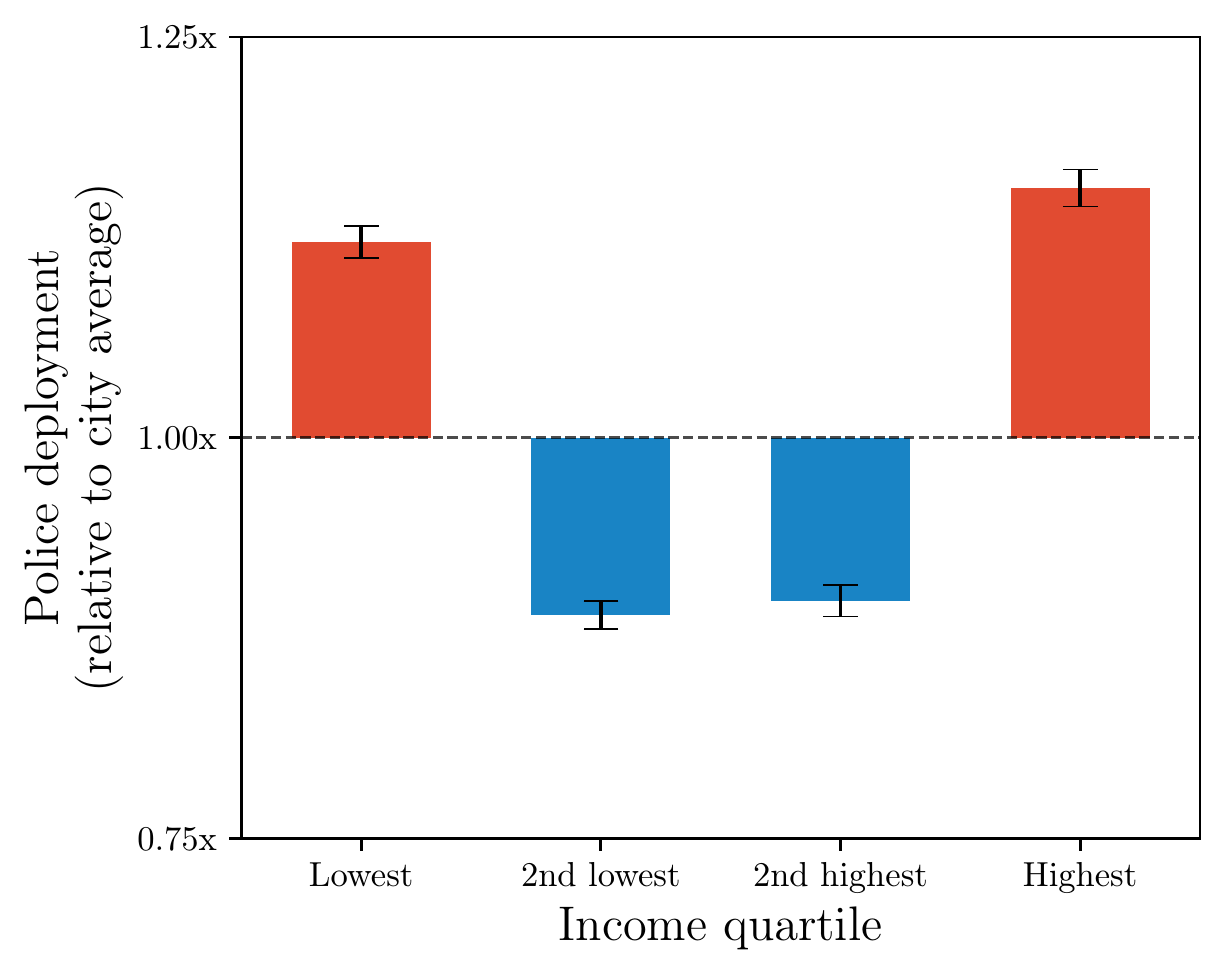}
         \caption{Deployments by income quartile.}
         \label{fig:income}
     \end{subfigure}
    \caption{Disparities in police deployments by race group (a) and median household income (b) without restricting to residential zones. Race group results are similar to the results restricting to residential zones. For income quartile, the relationship becomes U-shaped when not restricting to residential zones, with the highest police deployments in the top and bottom income quartiles.} 
    \label{fig:demographics-barplots-no-residential-restrictions}
\end{figure*}
\clearpage
\newpage

\section{Supplementary Tables}
\pagenumbering{gobble} 
\begin{table*}[b]
\center
\begin{tabular}{lrrrr}
\toprule
{} & \multicolumn{2}{c}{AUC} & \multicolumn{2}{c}{Average} \\
{} & \multicolumn{2}{c}{} & \multicolumn{2}{c}{Precision} \\
Above Median? & True  & False &             True  & False \\
Subgroup                                      &       &       &                   &       \\
\midrule
Manhattan                                     &  0.99 &  0.98 &              0.83 &  0.81 \\
Distance From Nearest Police Station > Median &  0.98 &  0.99 &              0.83 &  0.82 \\
Distance From Nearest Crime [6hr] > Median    &  0.99 &  0.99 &              0.78 &  0.84 \\
Distance From Nearest Crime [3hr] > Median    &  0.99 &  0.99 &              0.82 &  0.82 \\
Distance From Nearest Crime [1hr] > Median    &  0.99 &  0.99 &              0.82 &  0.82 \\
Population Density > Median                   &  0.99 &  0.98 &              0.82 &  0.81 \\
Median Household Income > Median              &  0.99 &  0.99 &              0.81 &  0.83 \\
Percent Hispanic > Median                     &  0.99 &  0.98 &              0.84 &  0.80 \\
Percent Asian > Median                        &  0.99 &  0.99 &              0.84 &  0.79 \\
Percent Black > Median                        &  0.99 &  0.98 &              0.86 &  0.77 \\
Percent White > Median                        &  0.99 &  0.99 &              0.81 &  0.82 \\
Daytime                                       &  0.98 &  0.99 &              0.81 &  0.82 \\
Weekend                                       &  0.98 &  0.99 &              0.77 &  0.83 \\
Phase 1                                       &  0.99 &  0.98 &              0.82 &  0.81 \\
\bottomrule
\end{tabular}

\vspace{6pt}
\caption{AUC (i.e., area under the ROC curve) and average precision by subgroup.}
\label{tab:subgroups}
\end{table*}

\begin{table*}[b]
\centering
\captionsetup{justification=centering}
    \begin{tabular}{lrrrr}
\toprule
{} &  Precision &  Recall &  AUC &   AP \\
\midrule
Validation Set &       0.78 &    0.83 & 0.98 & 0.82 \\
Test Set       &       0.81 &    0.77 & 0.99 & 0.82 \\
\bottomrule
\end{tabular}

    \vspace{6pt}
    \caption{Overall classifier performance statistics for the validation and test sets. \\Precision and recall are computed at a positive classification threshold of 0.77.} 
    \label{tab:test_set_performance_table}
\end{table*}

\end{document}